\documentclass[prx,twocolumn, superscriptaddress,
showpacs,
longbibliography,
aps,
showkeys]{revtex4-2}

\usepackage{dsfont} 
\usepackage{bm} 

\usepackage{bbold} 
\usepackage{xcolor}
\usepackage{soul}

\usepackage{float}
\usepackage{physics}
\usepackage{graphicx,braket}
\usepackage{amssymb} 

\usepackage{hyperref}
\hypersetup{colorlinks=true,linkcolor=blue,citecolor=blue}
\usepackage{amsmath,amssymb,amsfonts}
\usepackage{wrapfig}
\usepackage[export]{adjustbox}
\usepackage{cleveref} 
\usepackage{hhline}
\usepackage{siunitx}
\usepackage{booktabs}
 
\usepackage{enumitem}   

\usepackage{cleveref}
\crefname{equation}{Eqs.}{Eqs.}
\Crefname{equation}{Equation}{Equations}
\crefrangelabelformat{equation}{(#3#1#4--#5#2#6)}
\crefmultiformat{equation}{Eqs. (#2#1#3}{, #2#1#3)}{#2#1#3}{#2#1#3}
\Crefmultiformat{equation}{Equations (#2#1#3}{, #2#1#3)}{#2#1#3}{#2#1#3}

\usepackage{lipsum}

\begin{document}
	
	\allowdisplaybreaks 
	
	\flushbottom
	\title{Quantum metrology through spectral measurements in quantum optics}

	\author{Alejandro Vivas-Via{\~n}a}
	\email[]{vivasa@chalmers.se}
	\affiliation{Department of Microtechnology and Nanoscience, Chalmers University of Technology, 412 96 Gothenburg, Sweden}	
	\affiliation{Institute of Fundamental Physics IFF-CSIC, Calle Serrano 113b, 28006, Madrid, Spain}
	\author{Carlos S\'anchez Mu\~noz}
	\email[]{carlos.sanchez@iff.csic.es}
	\affiliation{Institute of Fundamental Physics IFF-CSIC, Calle Serrano 113b, 28006, Madrid, Spain}

	\newcommand{\down}{\op{g}{e}}
	\newcommand{\up}{\op{e}{g}}
	\newcommand{\downd}{\op{+}{-}} 
	\newcommand{\upd}{\op{+}{-}}
	\newcommand{\app}{a^\dagger}
	\newcommand{\ssp}{\sigma^\dagger}
	\newcommand*{\Resize}[2]{\resizebox{#1}{!}{$#2$}}%
	\newcommand{\conc}{\mathcal C(\rho)}
	
	\begin{abstract}
		Continuously monitored quantum systems are emerging as promising platforms for quantum metrology, where a central challenge is to identify measurement strategies that optimally extract information about unknown parameters encoded in the complex quantum state of emitted radiation.
		Different measurement strategies effectively access distinct temporal modes of the emitted field, and the resulting choice of mode can strongly impact the information available for parameter estimation.
		While a ubiquitous approach in quantum optics is to select frequency modes through spectral filtering, the metrological potential of this technique has not yet been systematically quantified.
		We develop a theoretical framework to assess this potential by modeling spectral detection as a cascaded quantum system, allowing us to reconstruct the full density matrix of frequency-filtered photonic modes and to compute their associated Fisher information.
		This framework provides a minimal yet general method to benchmark the performance of spectral measurements in quantum optics, allowing to identify optimal filtering strategies in terms of frequency selection, detector linewidth, and metrological gain accessible through higher-order frequency-resolved correlations and mean-field engineering.
		These results lay the groundwork for identifying and designing optimal sensing strategies in practical quantum-optical platforms.
	\end{abstract}

	\date{\today} \maketitle
	
	
	
	\section{Introduction }
	
	\begin{figure*}[t!]
		\centering
		\includegraphics[width=1.0\linewidth]{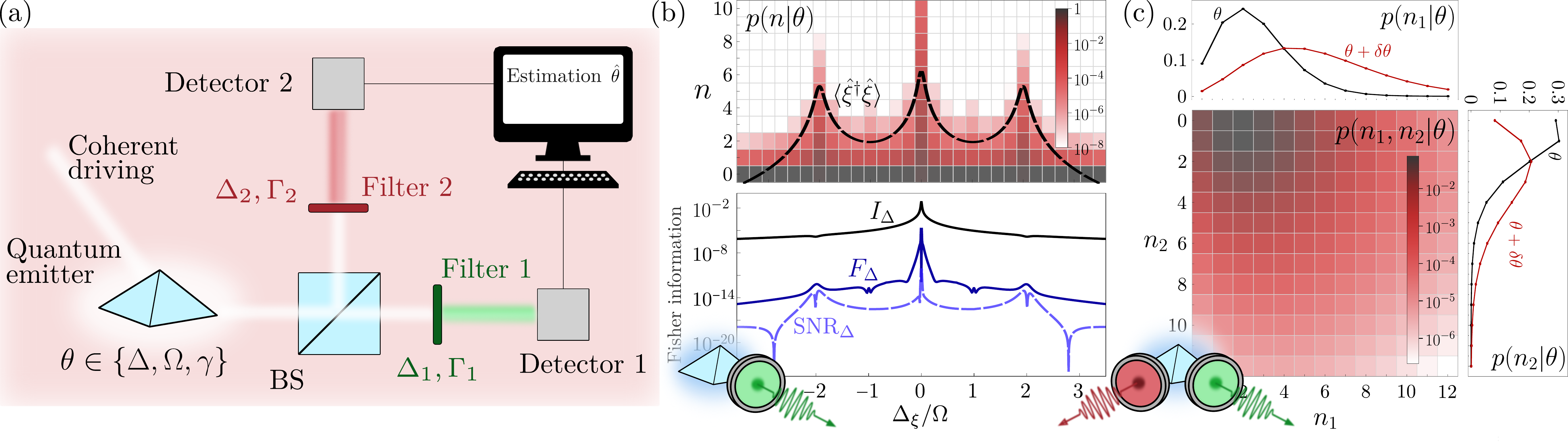}
		\caption{
			(a) Sketch of the metrological setup.
			A coherently driven emitter---characterized by the parameter to be estimated, $\theta\in\{\Delta,\Omega,\gamma\}$---emits radiation that is split by a beam splitter and routed into two frequency-selective sensors, each centered at $\Delta_i$ and featuring a spectral resolution $\Gamma_i$.
			(b) Metrological protocol. A sensor collects the radiation emitted by the source, giving access to the filtered density matrix of the output field in the frequency domain. This process allows the extraction of the $\theta$-dependent photon-number distribution, $p(n|\theta) = \langle n | \hat{\rho}_\theta | n \rangle$ (top panel). 
			An estimation strategy can then be employed to infer $\theta$ from $p(n|\theta)$. The precision of any unbiased estimator is bounded by the inverse of the Fisher information (bottom panel): 
			$I_\Delta$ (solid black), $F_\Delta$ (solid blue), and $\text{SNR}_\Delta$ (dashed blue), which gives the precision of an estimation based only on the mean population of the sensor $\langle \hat \xi^\dagger \hat \xi \rangle$ (black-dashed curve in the top panel) instead of the full photon-counting distribution. 
			All quantities correspond to the estimation of the qubit–laser detuning parameter $\Delta$.
			(c) Joint $\theta$-dependent probability distribution, $p(n_1, n_2 | \theta)$ for simultaneous detection with two sensors. The top and right panels display the marginal distributions for each detector, $p(n_i|\theta)$, highlighting their sensitivity to a small perturbation in the parameter, $\theta + \delta\theta$ (a non-infinitesimal perturbation $\delta\theta = -0.65\gamma$ was chosen to enhance the discernibility of the curves).
			Parameters:
			(b) $\Omega=10\gamma$, $\Delta=10^{-5}\gamma$, $\Gamma=10^{-1}\gamma$, $\varepsilon=0.9$.
			(c)
			$\Omega=2\gamma$, $\Delta=0$, $\Gamma=10^{-2}\gamma$, $\varepsilon=0.75$, $\Delta_1=0$, $\Delta_2=5\cdot10^{-3}\gamma$;
		}
		\label{fig:Fig1_setup}
	\end{figure*}
	Continuously monitored open quantum systems are emerging as promising platforms for quantum metrology~\cite{
		MabuchiInversionQuantum1996, 
		GambettaStateDynamical2001,
		TsangFundamentalQuantum2011, 
		TsangContinuousQuantum2012,
		GammelmarkBayesianParameter2013,
		WisemanQuantumMeasurement2014,
		DegenQuantumSensing2017,
		CortezRapidEstimation2017,
		AlbarelliUltimateLimits2017,
		GrossQubitModels2018,
		BraunQuantumenhancedMeasurements2018,
		ShankarContinuousRealTime2019,
		FallaniLearningFeedback2022,
		YangEfficientInformation2023,
		KhanahmadiQubitReadout2023,
		IliasCriticalityenhancedElectric2024,
		GoreckiInterplayTime2025,
		CabotQuantumEnhanced2025,
		GoreckiTimeCorrelations2025}, with key applications in areas such as atomic magnetometry~\cite{Amoros-BinefaNoisyAtomic2021, Amoros-BinefaNoisyAtomic2025}, force sensing~\cite{BarzanjehOptomechanicsQuantum2022}, gravitational wave detection~\cite{DanilishinQuantumMeasurement2012}, and criticality-enhanced metrology~\cite{
		MacieszczakDynamicalPhase2016,
		Fernandez-LorenzoQuantumSensing2017,
		IliasCriticalityEnhancedQuantum2022,
		DingEnhancedMetrology2022,
		DiCandiaCriticalParametric2023,
		PavlovQuantumMetrology2023,
		MontenegroQuantumMetrology2023,
		IliasCriticalityenhancedElectric2024,
		CabotExploitingNonequilibrium2024,
		CabotContinuousSensing2024,
		AlushiCollectiveQuantum2025,
		CabotQuantumEnhanced2025,
		MattesDesigningOpen2025}. These developments have been accompanied by significant theoretical progress, particularly in establishing a formal framework for quantum metrology in such systems. Notably, this includes the derivation of ultimate precision bounds based on the Cramér–Rao bound for information encoded in the radiated field~\cite{GammelmarkFisherInformation2014, CatanaFisherInformations2015, MacieszczakDynamicalPhase2016}.
	
	However, a major open challenge remains: devising measurement strategies that efficiently extract the full information content from the inherently multi-mode quantum state of the radiated field~\cite{YangEfficientInformation2023}. Standard time-local measurements, such as photon-counting or homodyne detection, fall short of saturating the quantum Fisher information (QFI)~\cite{GammelmarkFisherInformation2014,KiilerichParameterEstimation2015}, as they fail to capture information embedded in higher-order and non-time-local correlations.
	To overcome this limitation, recent strategies go beyond time-local measurements by incorporating ancillary systems---such as absorbers~\cite{GodleyAdaptiveMeasurement2023} or decoders~\cite{YangEfficientInformation2023}---that interact with and extract structured information from the radiated field. These approaches are closely tied to efforts aimed at reconstructing the full quantum state of the output field, often by coupling to ancillary modes designed to capture specific temporal profiles~\cite{KiilerichInputOutputTheory2019, KiilerichQuantumInteractions2020, YangEntanglementPhotonic2025}. This filtering of the output field naturally connects with frequency filtering, which can be understood as a physically motivated temporal-mode basis~\cite{EberlyTimedependentPhysical1977,HollandUnravelingQuantum1998} which, notably, can also be formally described in terms of ancillary modes~\cite{DelValleTheoryFrequencyFiltered2012,CarrenoExcitationQuantum2016,CarrenoExcitationQuantum2016a}. The expectation values obtained from the steady state of these ancillary modes correspond, by the ergodic hypothesis, to the time-averaged frequency-filtered signals continuously measured over a single realization~\cite{MolmerMonteCarlo1993,GardinerQuantumNoise2004,KummererPathwiseErgodic2004}.  Frequency filtering is routinely implemented in experimental setups via interferometers and spectrometers~\cite{UlhaqCascadedSinglephoton2012,PeirisTwocolorPhoton2015,SilvaColoredHanbury2016,PeirisFransonInterference2017}, providing a direct and accessible way to probe structured features of the radiated field.
	
	Despite this conceptual alignment, frequency-resolved measurements have yet to be fully leveraged in the context of quantum metrology. Although multi-photon frequency correlations have been extensively studied---particularly in systems such as coherently driven two-level emitters~\cite{UlhaqCascadedSinglephoton2012, PeirisTwocolorPhoton2015, Gonzalez-TudelaTwophotonSpectra2013, SanchezMunozViolationClassical2014, LopezCarrenoPhotonCorrelations2017, ZubizarretaCasalenguaConventionalUnconventional2020, LopezCarrenoEntanglementResonance2024, YangEntanglementPhotonic2025}, cavity QED~\cite{Gonzalez-TudelaTwophotonSpectra2013}, optomechanical~\cite{SchmidtFrequencyresolvedPhoton2021} and molecular platforms~\cite{Martinez-GarciaCoherentElectronVibron2024, MoradiKalardePhotonAntibunching2025}---their potential as a resource for parameter estimation and precision sensing remains largely untapped.
	This represents a significant opportunity, as the light emitted by continuously driven quantum systems displays rich structures in their frequency-frequency correlation functions, which are expected to encode valuable information about the underlying system parameters.
	
	In this work, we establish a theoretical framework to quantify the metrological potential of frequency-resolved measurements in quantum optics. This enables us to address key questions, such as:
	(i) Can frequency filtering enhance the amount of information extractable from standard quantum optical measurements?
	(ii) Does accounting for quantum fluctuations lead to improved sensitivity?
	(iii) Do quantum correlations between different frequency components contribute to increased metrological precision?

	In most of this work, we apply our formalism to a minimal yet highly instructive system: a single two-level emitter coherently driven near resonance. Despite its simplicity, this setup exhibits considerable complexity and reveals a rich landscape of multi-photon processes~\cite{DelValleTheoryFrequencyFiltered2012, UlhaqCascadedSinglephoton2012, PeirisTwocolorPhoton2015, Gonzalez-TudelaTwophotonSpectra2013, SanchezMunozViolationClassical2014, LopezCarrenoPhotonCorrelations2017, ZubizarretaCasalenguaConventionalUnconventional2020, LopezCarrenoEntanglementResonance2024, YangEntanglementPhotonic2025}. It serves as an ideal testbed for exploring the emergence of non-classical correlations in the emitted light and for addressing some of the key questions outlined above, i.e., exploring the role of such correlations in enhancing quantum parameter estimation of unknown atomic parameters. 
	%
	%
	%
	%
	%
	%
	%
	%
	We assess the metrological potential of frequency filtering by evaluating the classical Fisher information associated with coherently displaced photon-counting measurements of frequency filtered modes. Specifically, we quantify the sensitivity of estimating an unknown atomic parameter based on photons detected at a given frequency $\omega$, with spectral resolution characterized by a linewidth $\Gamma$, and analyze how this estimation is enhanced when frequency correlations between photons detected at different frequencies are taken into account.

	The manuscript is organized as follows.
	In Section~\ref{sec:sec2}, we introduce the model and outline the metrological strategy.
	In Section~\ref{sec:sec3}, we focus on the single-sensor case, where we define the frequency-resolved Fisher information and analyze its key features, including its spectral dependence and sensitivity to the filtering process. We also present a mean-field engineering approach to optimize the Fisher information.
	In Section~\ref{sec:sec4}, we extend the framework to two sensors, demonstrating how photon-photon correlations can be exploited to improve parameter estimation.
	Finally, in Section~\ref{sec:Generality}, we illustrate the generality of the framework by analyzing its application to a transmon qubit and an optomechanical system.

	\section{General framework}
	\label{sec:sec2}
	
	In this section, we introduce the physical model and the metrological strategy for estimating unknown atomic parameters. While the framework is fully general and applicable to a broad class of systems, we focus here on a minimal implementation sketched in Fig.~\ref{fig:Fig1_setup}(a): frequency-filtering of the radiation emitted by a coherently driven quantum emitter. This setting allows us to clearly illustrate the core features and advantages of the proposed method.
	
	\subsection{Model}
	
	We describe the quantum source as a coherently driven two-level system (TLS)---which we will also refer to as a qubit---, spanning a basis $\{|g\rangle, |e\rangle \}$, which denotes the ground and excited states, respectively, such that the lowering operator gets defined as $\hat \sigma\equiv |g\rangle \langle e|$. 
	The quantum emitter is driven by a coherent field characterized by a Rabi frequency $\Omega$ and a laser frequency $\omega_L$. In the rotating frame of the laser, and under the rotating wave approximation, the Hamiltonian of the source reads
	\begin{equation}
		\hat H= \Delta \hat \sigma^\dagger \hat \sigma + \Omega(\hat \sigma  + \hat \sigma^\dagger),
		\label{eq:TLS_Hamiltonian}
	\end{equation}
	where $\Delta \equiv \omega_\sigma- \omega_L$ is the qubit-laser detuning, with $\omega_\sigma$ being the natural frequency of the quantum emitter.
	The source undergoes incoherent processes, essentially de-excitation through spontaneous emission at a rate $\gamma$, resulting from its interaction with the vacuum electromagnetic field or a structured environment such as a coplanar waveguide~\cite{BreuerTheoryOpen2007, CarmichaelOpenSystems1993, GarciaRipollQuantumInformation2022}. In the Markovian regime, the dissipative dynamics of the TLS---described by its reduced density matrix $\hat{\rho}_\mathrm{q}$---, is governed by the following master equation:
	\begin{equation}
		\frac{d \hat \rho_\mathrm{q} }{dt}= -i [\hat H, \hat \rho_\mathrm{q}] + \frac{\gamma}{2} \mathcal{D}[\hat \sigma] \hat \rho_\mathrm{q},
		\label{eq:MasterEqTLS}
	\end{equation}
	where $\mathcal{D}[\hat \sigma]$ is the Lindblad superoperator, defined as $\mathcal{D}[\hat X] (\cdot)\equiv 2\hat X (\cdot) \hat X^\dagger - \{ \hat X^\dagger \hat X,(\cdot) \} $. 
	Considering now the output field $\hat a_\text{out}(t)$ radiated by the source, we can define any temporal mode with bosonic statistics as $\hat a_v = \int_{-\infty}^{\infty} v^*(t) \hat a_\text{out}(t) \, dt$, provided a normalized filter function, $\int_{-\infty}^{\infty} |v(t)|^2\, dt = 1$.
	In order to describe frequency-filtered modes of the emitted field and obtain their full quantum state, we will recourse to the time-independent sensor method~\cite{DelValleTheoryFrequencyFiltered2012,CarrenoExcitationQuantum2016,CarrenoExcitationQuantum2016a}, which, in the case of frequency-filtering, is completely equivalent to the time-dependent input-output theory of quantum pulses developed in Refs.~\cite{KiilerichInputOutputTheory2019,KiilerichQuantumInteractions2020}, as we show below.
	In the latter approach, the system and the emitted radiation are modeled as a cascaded quantum network~\cite{GardinerQuantumNoise2004,CombesSLHFramework2017}, where the desired temporal mode $\hat a_v(t)$ is described by an ancillary cavity (with annihilation operator $\hat \xi$), driven non-reciprocally by the system with a time-dependent coupling rate
	\begin{equation}
		g_v(t)=-\frac{v^*(t)}{\sqrt{\int_0^tdt' |v(t)|^2}}.
		\label{eq:TimeDependentCoupling}
	\end{equation}
	It can be shown~\cite{KiilerichInputOutputTheory2019,KiilerichQuantumInteractions2020} that this choice of $g_v(t)$ results in $\hat\xi = \hat a_v$, meaning that the state of the ancillary sensor matches the state of the selected mode of the radiated field.   
	To describe the particular case of frequency-filtering, we set the temporal mode $v(t)$ as the normalized Lorentzian band-pass filter~\cite{EberlyTimedependentPhysical1977},
	\begin{equation}
		v(t)=\sqrt{\Gamma}e^{(t_0-t)(i\omega_\xi-\Gamma/2)}e^{i\phi}\Theta(t_0-t),
		\label{eq:LorentztianFilterFun}
	\end{equation}
	where $\omega_\xi$ is the frequency of the filtered mode, $\Gamma$ is the linewidth of the Lorentzian filter, $\phi$ is a constant phase, and $\Theta(t)$ is the Heaviside step function.
	This filter function describes the time-dependent physical spectrum of light~\cite{EberlyTimedependentPhysical1977}, with the time-dependence encoded in the free parameter $t_0$. Since Eq.~\eqref{eq:TimeDependentCoupling} assumes $t=0$ as the starting point of the dynamics (with the ancillary modes in vacuum), $v(t)$ is a well-normalized function in the steady-state limit $t_0\rightarrow \infty$. In that limit, the time-dependent coupling (in the rotating frame of the drive) simplifies to
	\begin{equation}
		g_v(t)=-\sqrt{\Gamma}e^{i (\Delta_\xi t +\phi)},        \label{eq:TimeDependentCoupling_result}
	\end{equation}
	where $\Delta_\xi\equiv \omega_\xi- \omega_L$ is the sensor-laser detuning. The trivial time-dependence of this coupling can be removed by incorporating a Hamiltonian term $\Delta_\xi \hat \xi^\dagger \hat \xi$, which yields a simple time-independent coupling $g_v =\sqrt\Gamma$ (upon setting $\phi=\pi$), in direct correspondence to the sensor approach from Refs.~\cite{DelValleTheoryFrequencyFiltered2012,CarrenoExcitationQuantum2016,CarrenoExcitationQuantum2016a}. 

	Consequently, the cascaded master equation governing the joint dynamics of the emitter and the ancillary sensor cavity---described by their reduced density matrix, $\hat \rho_{\mathrm{q,1}}$--- is given by [see details in Appendix~\ref{Appendix:Derivation-MasterEq}]
	\begin{multline}
		\frac{d \hat \rho_{\mathrm{q,1}} }{dt}= -i [\hat H+ \Delta_\xi \hat \xi^\dagger \hat \xi, \hat \rho_{\mathrm{q,1}}] + \frac{\gamma}{2} \mathcal{D}[\hat \sigma] \hat \rho_{\mathrm{q,1}} 
		\\
		+  \frac{\Gamma}{2} \mathcal{D}[\hat \xi]\hat \rho_{\mathrm{q,1}} 
		-
		\sqrt{\varepsilon \gamma \Gamma} \left\{ [\hat \xi^\dagger, \hat \sigma  \hat \rho_{\mathrm{q,1}} ] + [\hat \rho_{\mathrm{q,1}} \hat \sigma^\dagger, \hat \xi] \right\}.
		\label{eq:cascadedmaster1}
	\end{multline}
	Additionally, we have introduced an imperfect coupling factor $\varepsilon\leq 1$, accounting, e.g., for possible losses in the transmission from the source to the target~\cite{GardinerQuantumNoise2004} or to retain
	the scattered light from a coherently driven system~\cite{CarrenoExcitationQuantum2016,CarrenoExcitationQuantum2016a}. Therefore, $\varepsilon=1$ corresponds to perfect cascaded coupling, and $\varepsilon=0$ to the limit of no coupling. 

	Although the single-sensor case is relatively straightforward to describe mathematically, the extension to two or more cascaded sensors requires additional attention, since there are multiple descriptions which correspond to different physical implementations
	of the multi-mode measurement. 
	In this work, we will explore the specific case in which,
	prior to the filtering of each mode, the output signal is physically split~\cite{LopezCarrenoFrequencyresolvedMonte2018,LopezCarrenoEntanglementResonance2024} (e.g., by
	a balanced beam splitter in a Hanbury-Brown Twiss setup), as illustrated in Fig.~\ref{fig:Fig1_setup}(a). 
	The resulting dynamics of the quantum emitter and the two frequency-filtered modes is described by the joint density matrix $\hat\rho_{\mathrm{q,2}}$, whose evolution obeys the cascaded master equation [see details in Appendix~\ref{Appendix:Derivation-MasterEq}]
	\begin{multline}
		\frac{d \hat \rho_{\mathrm{q,2}} }{dt}= -i [\hat H+ \sum_{i=1}^2\Delta_{i} \hat \xi_i^\dagger \hat \xi_i, \hat \rho_{\mathrm{q,2}}] + \frac{\gamma}{2} \mathcal{D}[\hat \sigma] \hat \rho _{\mathrm{q,2}} 
		\\
		+ \sum_{i=1}^2 \frac{\Gamma_{i}}{2} \mathcal{D}[\hat \xi_i]\hat \rho_{\mathrm{q,2}}	-
		\sqrt{\varepsilon \gamma \Gamma_{i}/2} \left\{ [\hat \xi_i^\dagger, \hat \sigma  \hat \rho_{\mathrm{q,2}} ] + [\hat \rho_{\mathrm{q,2}} \hat \sigma^\dagger, \hat \xi_i] \right\}.
		\label{eq:cascadedmaster2}
	\end{multline}
	In both cascaded master equations, Eq~\eqref{eq:cascadedmaster1} and Eq.~\eqref{eq:cascadedmaster2}, $\hat{H}$ denotes the Hamiltonian of the quantum source as given in Eq.~\eqref{eq:TLS_Hamiltonian}. 
	The operator $\hat \xi_i$ represents the $i$th bosonic mode associated with each sensor, characterized by a sensor-laser detuning $\Delta_{i}\equiv \omega_{i}-\omega_L$, and linewidth $\Gamma_{i}$. 
	To relax the notation in the two-sensor case, we have replaced the variable $\Delta_\xi \rightarrow \Delta_i$ ($i=1,2$).
	Throughout this work we consider identical sensor linewidths, $\Gamma_{1}=\Gamma_{2}=\Gamma$, and thus identical cascaded coupling strengths due to Eq.~\eqref{eq:TimeDependentCoupling_result}.
	The factor $1/\sqrt{2}$ in Eq.~\eqref{eq:cascadedmaster2} accounts for vacuum contributions arising from signal splitting at the balanced beam splitter.
	In contrast, in a previous work~\cite{YangEntanglementPhotonic2025}, we considered the situation in a waveguide setup in which the photonic modes are digitally filtered by post-processing the signal,  resulting in a different master equation from Eq.~\eqref{eq:cascadedmaster2} that does not include the factor $1/\sqrt{2}$ in the cascaded-coupling term. 
	As we mentioned, extensions to $N\geq2$ sensors would require additional attention. These could be implemented using multiport beam-splitter, as in multi-copy quantum information protocols~\cite{IzumiQuantumReceivers2013,NotarnicolaOptimizingStatediscrimination2023,NotarnicolaBeatingStandard2023}, by employing cascaded cavity arrays~\cite{KiilerichInputOutputTheory2019,KiilerichQuantumInteractions2020} or frequency-dispersed photon detection schemes~\cite{SilvaColoredHanbury2016}. 
	

	\subsection{Metrological strategy}
	
	We assess the metrological potential of the emitted light within the framework of quantum parameter estimation~\cite{DowlingQuantumOptical2015,LiuQuantumFisher2020,ParisQUANTUMESTIMATION2009,PetzIntroductionQuantum2011,WisemanQuantumMeasurement2009,Demkowicz-DobrzanskiQuantumLimits2015,PezzeQuantumMetrology2018,PolinoPhotonicQuantum2020,BarbieriOpticalQuantum2022}, specifically by introducing the $N$-sensor frequency-resolved classical Fisher information (CFI) obtained from photon-counting measurements.
	To allow tunable measurement strategies, we consider displaced photon-counting measurements, where the photonic quantum state $\hat \rho_\theta$---encoding the parameter $\theta$ to be estimated---is coherently displaced prior to detection.
	Therefore, the resulting CFI, $F^{\vec\alpha}_\theta$, takes the form
	\begin{equation}
		F^{\vec\alpha}_\theta= \sum_{\vec{n}}^\infty \frac{1}{p^{\vec \alpha}(\vec{n}|\theta)} \left(\frac{\partial p^{\vec \alpha}(\vec{n}|\theta)}{\partial \theta}\right)^2,
		\label{eq:CFI_PC}
	\end{equation}
	with the joint conditional probability distribution 
	\begin{equation}
		p^{\vec \alpha}(\vec{n}|\theta)=\langle \vec{n}|\hat D(\vec \alpha) \hat \rho_\theta\hat D^\dagger(\vec  \alpha)| \vec{n}\rangle,
		\label{eq:ConditionedProbDistrb}
	\end{equation}
	where $\hat D(\vec \alpha)=\prod_{i=1}^N\hat D(\alpha_i)$ is the multi-mode displacement operator, with each $\alpha_i \in \mathbb{C}$ corresponding to the coherent displacement applied to the $i$th sensor, and $|\vec n\rangle\equiv |n_1,\ldots,n_N\rangle$, with $|n_i\rangle$ denoting the photon number eigenstates of the $i$th sensor.
	We implement this strategy by performing $N$ discrete frequency-resolved displaced photon-counting measurements, with each frequency channel modeled as an ancillary single-mode bosonic cavity. We emphasize that the number of detection channels $N$ corresponds to the number of modeled frequency filters---sensors---, not to repeated experimental realizations.
	Formally, the measurement process is described by the tensor product of sets of operators describing the POVMs, $ \hat \Lambda=\{\bigotimes_{i=1}^N\hat \Lambda_{n,\alpha_i}^{(i)}\}$,
	where 
	\begin{equation}
		\hat \Lambda_{n,\alpha_i}^{(i)} =\hat D^\dagger(\alpha_i)|n_i \rangle \langle n_i |\hat D
		(\alpha_i).
		\label{eq:POVMs}
	\end{equation}
	Physically, displaced photon-counting can be realized by mixing the output field of the quantum emitter with a local oscillator $\alpha(t)$ via a highly transmissive beam splitter, yielding an effective detection signal $\sqrt{\gamma} \hat{\sigma}(t) + \alpha(t)$. This mean-field engineering technique~\cite{ZubizarretaCasalenguaTuningPhoton2020,KimUnlockingMultiphoton2025,BrachtTunableMultiphoton2025} enables us to tailor the photonic state to either isolate the quantum fluctuations or optimize the entire state for the estimation of specific atomic parameters. 
	Standard photon-counting corresponds to the special case $\vec \alpha =\vec 0$, denoted as $p(\vec{n}|\theta)\equiv p^{\vec \alpha=\vec 0}(\vec{n}|\theta)$ and $F_\theta \equiv F^{\vec \alpha=\vec0}_\theta$. 
	Although this formulation applies generally to any number of frequency-resolved sensors, we restrict our analysis to the one‐ and two‐sensor cases.
	Accordingly, we evaluate $p^\alpha(n|\theta)$ for $N=1$ 
	and $p^{\alpha_1,\alpha_2}(n_1,n_2|\theta)$ for $N=2$, as illustrated in Fig.~\ref{fig:Fig1_setup}(b) and (c), respectively.
	In this work, $\theta$ represents a Hamiltonian or a Lindblad parameter from the source master equation in Eq.~\eqref{eq:MasterEqTLS}, such as the qubit-laser detuning, the driving strength of the laser interacting with the quantum emitter, or the emitter decay rate, $\theta \in \{\Delta, \Omega, \gamma \}$.
	In situations where only the mean value of an observable $\hat{O}$ is experimentally accessible---e.g., the mean photon number of the sensor---, we can quantify the estimation performance via the signal-to-noise ratio (SNR)~\cite{TothQuantumMetrology2014,MacieszczakDynamicalPhase2016,RamsLimitsCriticalityBased2018,DiCandiaCriticalParametric2023}
	\begin{equation}
		\text{SNR}_\theta[\hat O]=\frac{[\partial_\theta\langle\hat O \rangle]^2}{\langle \Delta^2\hat O\rangle},
		\label{eq:SNR}
	\end{equation}
	where $\langle \Delta^2\hat O\rangle=\langle\hat O^2\rangle-\langle \hat O\rangle^2$ is the variance, satisfying the relation $F_\theta^{\vec \alpha}\geq \text{SNR}_\theta[\hat O]$. 
	Essentially, the CFI quantifies, on average, the sensitivity of the conditional probability distribution to small changes on the parameter to estimate [see top and right panels of Fig.~\ref{fig:Fig1_setup}(b)], whereas the SNR quantifies the sensitivity based solely on changes in the observable mean value.
	
	The fundamental limit to the precision achievable by any unbiased estimator $\hat{\theta}$ is set by the quantum Cramér-Rao bound (QCRB)~\cite{CramerMathematicalMethods1991,RaoInformationAccuracy1992,PolinoPhotonicQuantum2020},
	\begin{equation}
		\Delta^2 \theta \geq \frac{1}{M  \text{SNR}_\theta} \geq \frac{1}{M F^{\vec{\alpha}}_\theta} \geq \frac{1}{M I_\theta}.
		\label{eq:QCRB}
	\end{equation}
	where $M$ is the number of repetitions, and $I_\theta$ denotes the quantum Fisher information (QFI), defined as the optimization of the CFI over all POVMs, or equivalently, the optimization of the SNR over all observables~\cite{GoreckiInterplayTime2025},  $I_\theta=\text{max}_{\{\hat \Lambda\}}F_\theta=\text{max}_{\{\hat O\}}\text{SNR}_\theta$. 
	For a general mixed state, the QFI can be expressed in closed form as~\cite{ParisQUANTUMESTIMATION2009}
	\begin{equation}
		I_\theta= \sum_{n,m}\frac{2}{\rho_n+\rho_m}|\langle \psi_n|\partial_\theta \hat \rho_\theta|\psi_m\rangle|^2,
	\end{equation}
	where $\hat{\rho}_\theta=\sum_n \rho_n  |\psi_n\rangle \langle \psi_n |$ is the eigen-decomposition of the quantum state.
	Throughout this work, we use these three figures of merit---QFI, CFI, and SNR---to characterize the metrological performance of our proposed strategy.
	We finally give here a relevant caveat regarding the QFI. Here, the QFI is computed over the quantum state of the sensors, which captures the filtered version of the emitted light. While it bounds the precision achievable with optimal measurements on the sensor modes, it does not necessarily correspond to the absolute lower bound for $\Delta^2 \hat \theta$, since it excludes information potentially available in the remaning unmeasured modes of the total radiation of the emitter. This absolute value is quantified by general expressions of the QFI as developed in Refs.~\cite{GammelmarkFisherInformation2014,MacieszczakDynamicalPhase2016}.

	Figure~\ref{fig:Fig1_setup}(b) illustrates the application of our framework for the single-sensor case. 
	The radiation emitted by the source is filtered by the sensor, yielding a frequency-resolved density matrix from which the conditional photon-count distribution $p(n|\theta)$ is extracted [top panel]. 
	In this example, the distribution, obtained via standard photon-counting measurements ($\alpha=0$), forms the basis for estimating the parameter $\theta$ with a precision set by the CFI, shown by the blue solid curve. 
	The example highlights the advantage of using the full photon-count statistics over relying only on the mean photon number. This is evident from the comparison between the CFI and the SNR of the sensor photon number (blue dashed curve). Both quantities lie below the QFI (black solid curve), indicating that neither corresponds to the optimal measurement for estimating the detuning ($\theta = \Delta$).
	As an illustration of the extension to the two-sensor case, Fig.~\ref{fig:Fig1_setup}(c) displays the joint distribution $p(n_1, n_2 | \theta)$ for $\alpha=0$ for the detection at two different frequencies, revealing off-diagonal structure that signals cross-correlations between detection channels, a feature that can be harnessed to enhance parameter estimation.

	\section{Frequency-resolved Fisher Information}
	\label{sec:sec3}
	
	In this section, we show in detail the application of the proposed metrological strategy based on frequency-resolved spectral measurements to the particular case of a single sensor to estimate unknown atomic parameters of the quantum source.
	Specifically, we compute the stationary CFI from displaced photon-counting measurements via Eq.~\eqref{eq:CFI_PC}, evaluated at a single frequency. We do so by solving for the steady-state density matrix $\rho_\theta^{\text{ss}}$ and its differentiation over the parameter to be estimated $\partial_\theta\rho_\theta^{\text{ss}}$ for the single-sensor cascaded master equation in Eq.~\eqref{eq:cascadedmaster1} [see Appendix~\ref{Appendix:method_SS} for computational details].
	Although the analysis of the results is general for any atomic parameter, in this section, we focus on estimating the qubit-laser detuning, $\theta=\Delta$. Results for other parameters, including the driving strength $\Omega$ and the decay rate $\gamma$, are provided in Appendix~\ref{Appendix:One_sensor_Other_parameters}.
	\begin{figure}[b!]
		\centering
		\includegraphics[width=1.\linewidth]{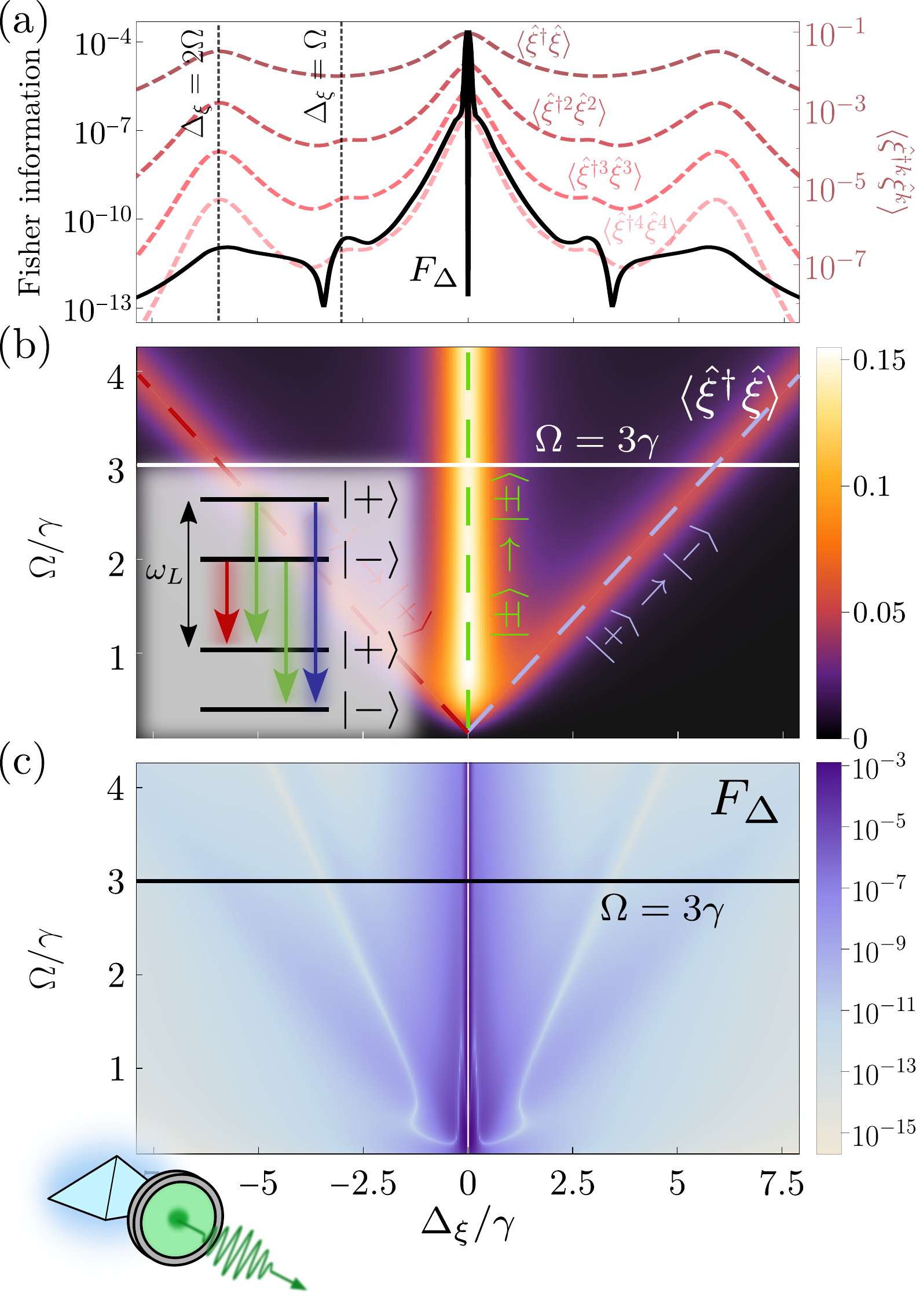}
		\caption{
			Frequency-resolved Fisher information for one-sensor for the estimation of $\theta=\Delta$.
			(a) $F_\Delta$ (black solid curve) and $\langle \hat \xi^{\dagger k} \hat \xi^k \rangle$ (red dashed curves, $k=1,2,3,4$)  in the frequency domain for a fixed value of $\Omega=3\gamma$.
			(b) $\langle \hat \xi^\dagger \hat \xi \rangle$ (fluorescence spectrum of the source) and (c) $F_\Delta$ in terms of the driving strength $\Omega$ and the detection frequency $\Delta_\xi$. The Mollow ladder is illustrated in the inset of panel (b).
			Parameters: (a) $\Omega=3\gamma$; (a-c) $\Delta=10^{-6}\gamma$, $\Gamma=10^{-1}\gamma$, $\varepsilon=0.1$.
		}
		\label{fig:Single-sensor-Fisher-DrivingFrequency}
	\end{figure}

	\subsection{Spectral dependence of the Fisher information}
	
	As a first step of our analysis, we study the CFI of the signal, $F_\Delta$,  using Eq.~\eqref{eq:CFI_PC} [setting $\alpha=0$ in Eq.~\eqref{eq:ConditionedProbDistrb}], and demonstrate how access to the spectral content of the emission enables the identification of optimal detection frequencies that enhance estimation sensitivity.
	Near resonance, $\Delta\approx0$ ($\omega_L\approx\omega_\sigma$), the emitter exhibits resonance fluorescence~\cite{LodahlInterfacingSingle2015,FlaggResonantlyDriven2009,KimbleTheoryResonance1976,AstafievResonanceFluorescence2010}, a cornerstone of quantum optics and a source of perfect antibunched photons. 
	The spectral features of the emitted light depend strongly on the excitation intensity $\Omega$ with respect the emission rate $\gamma$, giving rise to two regimes, as illustrated in Figs.~\ref{fig:Single-sensor-Fisher-DrivingFrequency}(a,b): the Heitler regime ($\Omega < \gamma/8$)~\cite{HeitlerQuantumTheory1984}, dominated by the Rayleigh-scattered light, and the Mollow regime ($\Omega>\gamma/8$)~\cite{MollowPowerSpectrum1969}, dominated by the inelastic-scattered light.
	In this regime, the energy levels of the TLS become dressed by the laser field, forming an infinite ladder of excitation manifolds comprised of dressed states~\cite{JaynesComparisonQuantum1963,Cohen-TannoudjiAtomPhotonInteractions1998, DalibardCorrelationSignals1983}, $|\pm\rangle$, formed by a superposition of the bare states of the TLS-laser system, and energy split by
	\begin{equation}
		\omega_S\equiv2\sqrt{\Omega^2+(\Delta/2)^2}.
	\end{equation}
	Energy transitions between adjacent excitation manifolds yields the characteristic Mollow triplet~\cite{MollowPowerSpectrum1969} [see Fig.~\ref{fig:Single-sensor-Fisher-DrivingFrequency}(a,b)]: the central peak corresponds to a doubly-degenerate transition, $|\pm \rangle \rightarrow |\pm\rangle$ ($\omega=\omega_L$), while the two side peaks correspond to transitions between different eigenstates, $|\pm \rangle \rightarrow |\mp \rangle$ $ (\omega=\omega_L \pm \omega_S$).

	A similar structure emerges in the frequency-resolved CFI, as shown in Figs.~\ref{fig:Single-sensor-Fisher-DrivingFrequency}(a) and~\ref{fig:Single-sensor-Fisher-DrivingFrequency}(c), where the stationary value of $F_\Delta$ is plotted as a function of both the filter frequency $\Delta_\xi$ and the driving strength $\Omega$, respectively. 
	The CFI features rich spectral patterns that reflect the structure of the resonance fluorescence spectrum~\cite{HeitlerQuantumTheory1984,MollowPowerSpectrum1969}. This results highlight the importance of selectively resolving the spectral components of radiation in optimizing metrological performance. This is the case even in the presence of severely imperfect coupling conditions [see Appendix~\ref{Appendix:ImperfectCoupling}]. 

	In Fig.~\ref{fig:Single-sensor-Fisher-DrivingFrequency}(a), we show a frequency slice of $F_\Delta$ for $\Omega = 3\gamma$ (black solid curve), revealing a structure reminiscent of the Mollow triplet. 
	This behavior reflects the underlying frequency-dependent features of the resonance fluorescence spectrum, characterized by the steady-state sensor population~\cite{DelValleTheoryFrequencyFiltered2012,CarrenoExcitationQuantum2016,CarrenoExcitationQuantum2016a}, $\langle \hat \xi^\dagger \hat \xi\rangle\sim \text{Re}\int_{0}^\infty d\tau e^{(i\Delta_\xi-\Gamma/2) \tau}\langle\hat \sigma^\dagger(0)\hat \sigma(\tau)\rangle$.
	Prominent features in $F_\Delta$ appear near the resonant frequencies of the Mollow triplet: $F_\Delta\rightarrow0$ ($\Delta_\xi \approx 0$)  and two local maxima ($\pm 2\Omega$). 
	These features indicate a well-defined spectral structure in terms of the sensor-laser detuning $\Delta_\xi$, revealing frequency windows where the estimation protocol is either significantly enhanced or strongly suppressed.
	However, additional nontrivial features emerge that are not captured by the first-order coherence alone, such as, the local minima and maxima near $\Delta_\xi \approx \pm \Omega$.
	In fact, these features are more closely related to higher-order moments, $\langle \hat \xi^{\dagger k}\hat \xi^k\rangle$ with $k\geq 2$, which reflect multi-photon processes beyond the fluorescence spectrum.
	This connection becomes evident when comparing $F_\Delta$ to higher-order correlators, such as $\langle \hat \xi^{\dagger 2}\hat \xi^2 \rangle$, $\langle \hat \xi^{\dagger 3}\hat \xi^3 \rangle$, and $\langle \hat \xi^{\dagger 4}\hat \xi^4 \rangle$, which show qualitative similarities with the additional structure in the Fisher information, as shown in the red dashed curves in Fig.~\ref{fig:Single-sensor-Fisher-DrivingFrequency}(a).

	These observations highlight a central point: although the CFI in Eq.~\eqref{eq:CFI_PC} depends solely on the diagonal elements of the sensor density matrix, it cannot be straightforwardly interpreted in terms of standard optical observables.
	The underlying reason is that the photon-counting probability $p(n|\theta)$ [see Eq.~\eqref{eq:ConditionedProbDistrb}] can be expressed as an infinite sum over normally ordered correlators~\cite{GardinerQuantumNoise2004,ZubizarretaCasalenguaStructureHarmonic2017},
	\begin{equation}
		p(n|\theta) = \sum_{k \geq n}^{n_{\text{exc}}} c_{n,k} \langle \hat{\xi}^{\dagger k} \hat{\xi}^k \rangle,
		\label{eq:single_prob_corr}
	\end{equation}
	where $c_{n,k}\equiv (-1)^{n+k}/n (k-n)!$, and $n_{\text{exc}}$ denotes the maximum number of excitations.
	Each term in the expansion captures the contribution of $k$-photon processes to the probability distribution. 
	As a result, the CFI---evaluated for an arbitrary coherent displacement $\alpha$---can be written as
	\begin{equation}
		F^\alpha_\theta=
		\sum_{n}^{n_{\text{exc}}} \frac{
			\left[ \sum_{k\geq n}^{n_{\text{exc}}} c_{n,k}  \partial_\theta \langle \hat D^\dagger(\alpha)\hat{\xi}^{\dagger k} \hat{\xi}^k\hat D(\alpha) \rangle \right]^2
		}{
			\sum_{k\geq n}^{n_{\text{exc}}}  c_{n,k}  \langle \hat D^\dagger(\alpha)\hat{\xi}^{\dagger k} \hat{\xi}^k\hat D(\alpha) \rangle
		}.
		\label{eq:FisherInform_Corr_OneSensor}
	\end{equation}
	This expression reveals that the frequency-resolved CFI is governed by a nontrivial interplay of the derivatives of all higher-order correlators with respect to the parameter $\theta$. 
	Consequently, well-known spectral signatures such as those in the mean photon number, $\langle \hat \xi^\dagger \hat \xi\rangle$, or second-order moment, $\langle \hat \xi^{\dagger 2} \hat \xi^2\rangle$, are generally insufficient to account for the full structure of the Fisher information, thereby explaining the appearance of features that have no direct analogue in conventional correlation functions.

	\subsection{Impact of sensor linewidth in parameter estimation}

	In our setup, each ancillary sensor acts as a frequency filter---see Eq.~\eqref{eq:LorentztianFilterFun}---, modeled by a Lorentzian band-pass filter function with linewidth $\Gamma$~\cite{DelValleTheoryFrequencyFiltered2012,CarrenoExcitationQuantum2016,CarrenoExcitationQuantum2016a}, providing a faithful physical description of photo-detection~\cite{EberlyTimedependentPhysical1977}.
	The linewidth $\Gamma$ sets the spectral ($\Delta\omega\sim \Gamma$) and temporal ($\Delta t\sim 1/\Gamma$) resolution of the measurement, as required by the uncertainty relations between energy and time~\cite{EberlyTimedependentPhysical1977,MandelstamUncertaintyRelation1991}. 
	The impact of filter linewidth on photonic statistics has been extensively studied in the context of a coherently driven qubit. In particular, the time--energy uncertainty introduced by filtering has a profound effect on frequency-resolved correlations~\cite{Gonzalez-TudelaTwophotonSpectra2013,SanchezMunozViolationClassical2014,LopezCarrenoPhotonCorrelations2017,LopezCarrenoJointSubnaturallinewidth2018,ZubizarretaCasalenguaConventionalUnconventional2020}. While color-blind measurements yield perfect antibunching, filtering introduces a finite time uncertainty that lifts this perfect single-photon character and reveals a rich landscape of frequency-resolved correlations. This non-trivial dependence extends to other quantities as well, including the degree of photonic entanglement~\cite{LopezCarrenoEntanglementResonance2024,YangEntanglementPhotonic2025}.
	Given its central role in the detection process and the properties defining the quantum photonic state, we now assess the metrological impact of the sensor linewidth.

\begin{figure}[t!]
		\centering
		\includegraphics[width=1.\linewidth]{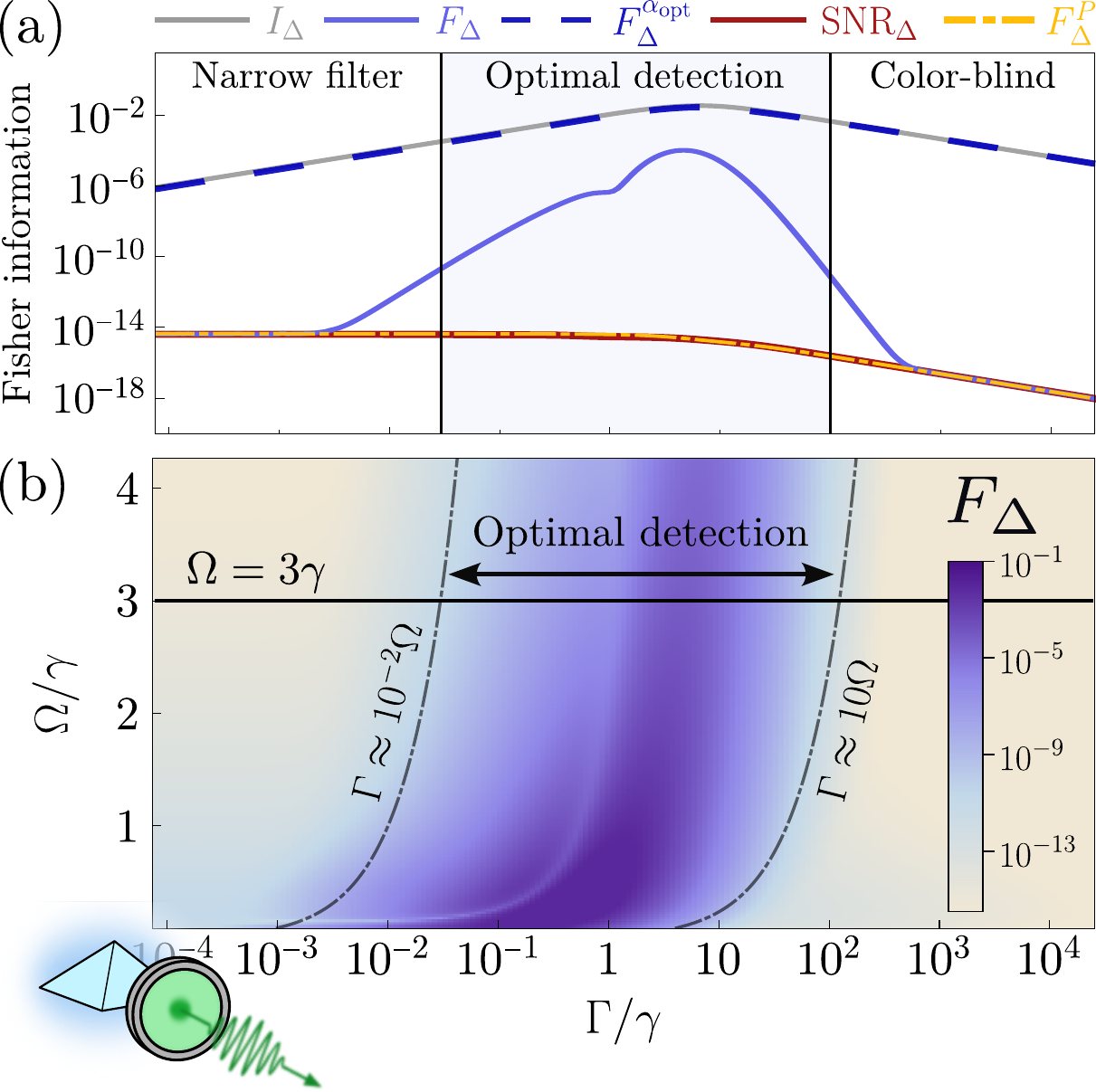}
		\caption{
			Impact of sensor linewidth $\Gamma$ on the performance of the estimation protocol for a single sensor. 
			(a) $I_\Delta$ (grey solid), $F_\Delta$ (blue solid), and $\text{SNR}_\Delta$ (red solid) as a function of $\Gamma$, shown for $\Omega=3\gamma$. 
			Additionally, $F_\Delta^{\alpha_{\text{opt}}}$ (blue dashed) and $F_\Delta^{P}$ (yellow dot-dashed) are shown.
			(b) $F_\Delta$ in terms of the driving strength $\Omega$ and the sensor linewidth $\Gamma$. The optimal regime is approximately bounded by $\Gamma \approx (10^{-2}\Omega,10\Omega)$, i.e., where the sensor preserves spectral resolution and metrological performance is enhanced. 
			Parameters: $\Delta=10^{-6}\gamma,\ \Delta_\xi=\Omega,\ \varepsilon=1$.
		}
		\label{fig:ImpactSensorLinewidth}
	\end{figure}

	Figure~\ref{fig:ImpactSensorLinewidth} shows the CFI as a function of the sensor linewidth $\Gamma$ and driving strength $\Omega$ for a fixed value of the sensor frequency, $\Delta_\xi=\Omega$. 
	Figure~\ref{fig:ImpactSensorLinewidth}(a) depicts the CFI in terms of $\Gamma$ for a particular choice of the driving strength ($\Omega=3\gamma$, blue solid).
	Although this value lies in the Mollow regime, we find that both the Heitler and Mollow regimes
	exhibit an optimal detection window that maximizes the CFI, occurring roughly in the range $\Gamma_{\text{opt}}\approx (10^{-2}\,\Omega,10\,\Omega)$. 
	Generally, in both extreme limits of the sensor linewidth---narrowband ($\Gamma \rightarrow 0$) and broadband ($\Gamma \rightarrow \infty$)---the CFI reaches minimal values.
	In the narrowband regime, the sensor achieves perfect spectral resolution but loses time definition, leading to a Gaussian description of the signal. As a result, when $\Gamma \ll 10^{-2} \Omega$, the CFI collapses onto the SNR (red curve), $F_\theta \approx \text{SNR}_\theta$, indicating that estimation is dominated by mean values.
	This behavior is consistent with previous reports observing frequency-resolved correlations at optimal ranges of filter bandwidth, narrow enough to resolve spectral features, but broad enough to limit the time uncertainty in the detection~\cite{Gonzalez-TudelaTwophotonSpectra2013,SanchezMunozViolationClassical2014,LopezCarrenoEntanglementResonance2024,YangEntanglementPhotonic2025}. 

	The comparison between $F_\theta$ and $\text{SNR}_\theta$ provides some intuition about the importance of quantum fluctuations in parameter estimation. An alternative measure for this assessment,  the Poissonian Fisher information (PFI), was introduced by the authors in Ref.~\cite{Vivas-VianaTwophotonResonance2021}. The PFI is based on a Poissonian approximation for the photon statistics~\cite{DelaubertQuantumLimits2008}, leading to an expression of the Fisher information given by [see Appendix~\ref{Appendix:Poissonian_Fisher}], 
	\begin{equation}
		F^P_\theta \equiv \frac{[\partial_\theta \langle \hat \xi^\dagger \hat \xi \rangle]^2}{ \langle \hat \xi^\dagger \hat \xi \rangle}.
		\label{eq:PoissonianFisher}
	\end{equation}
	Unlike the SNR, the PFI assumes coherent-state statistics, where the variance equals the mean. 
	In regimes where the signal is Gaussian but not fully coherent the this quantity may misestimate the accessible information.
	Nevertheless,
	$ F^P_\theta\approx \text{SNR}_\theta$ when $\langle\Delta^2 \hat \xi^\dagger\hat \xi\rangle \approx \langle  \hat \xi^\dagger\hat \xi\rangle$, as shown by the yellow dot-dashed curve in Fig.~\ref{fig:ImpactSensorLinewidth}(a). 
	Therefore, for faithful assessment of estimation performance when only the mean value is accessible, the SNR provides a more reliable benchmark than $F^P_\theta$, as it properly accounts for the actual statistical properties of the signal.

	In the broadband limit, the measurement becomes color-blind: temporal precision is perfectly defined, but spectral information is lost. 
	As a result, the CFI decreases and becomes equal to the SNR---and to the PFI---, signaling a reduction of the information contained in quantum fluctuations of the filtered field.

	These results showcase some of the nuances involved in the process of filtering radiation for the task of parameter estimation, as well as the potential improvements one can achieve by optimizing the process. When operating in the optimal detection regime [see Fig.~\ref{fig:ImpactSensorLinewidth}(b)], the sensor can boost the CFI by over ten orders of magnitude.
	Nonetheless, even in regimes where the CFI, SNR, and PFI get drastically suppressed---due to either extreme spectral or time resolution---, the quantum Fisher information (QFI) remains finite, revealing the existence of an optimal POVM different from the standard photon-counting measurement used here. Below, we discuss an update of the measurement scheme that can bring us closer to the limit set by the QFI.

	\subsection{Mean field engineering}

	Recent experimental works~\cite{KimUnlockingMultiphoton2025,BrachtTunableMultiphoton2025} showed that adjusting the weight of the coherent component of light via interference~\cite{VogelSqueezingAnomalous1991,BreitenbachMeasurementQuantum1997,SchulteQuadratureSqueezed2015,FischerSelfhomodyneMeasurement2016} serves as a novel control knob for engineering quantum statistics, termed mean-field engineering. This was verified by unlocking multiphoton effects at very low-driving in resonance fluorescence.
	%
	While the bare emission is antibunched, the engineered emission becomes superbunched at all orders when the classical field is perfectly suppressed. The nature of the multiphoton emission can thus be tuned by adjusting the amplitude of the interfering classical field~\cite{LopezCarrenoJointSubnaturallinewidth2018,ZubizarretaCasalenguaTuningPhoton2020,KimUnlockingMultiphoton2025,BrachtTunableMultiphoton2025}. 
	The potential of this mean-field engineering technique in quantum metrology and frequency-resolved detection is still unexplored. 

\begin{figure}[t!]
		\centering
		\includegraphics[width=1.\linewidth]{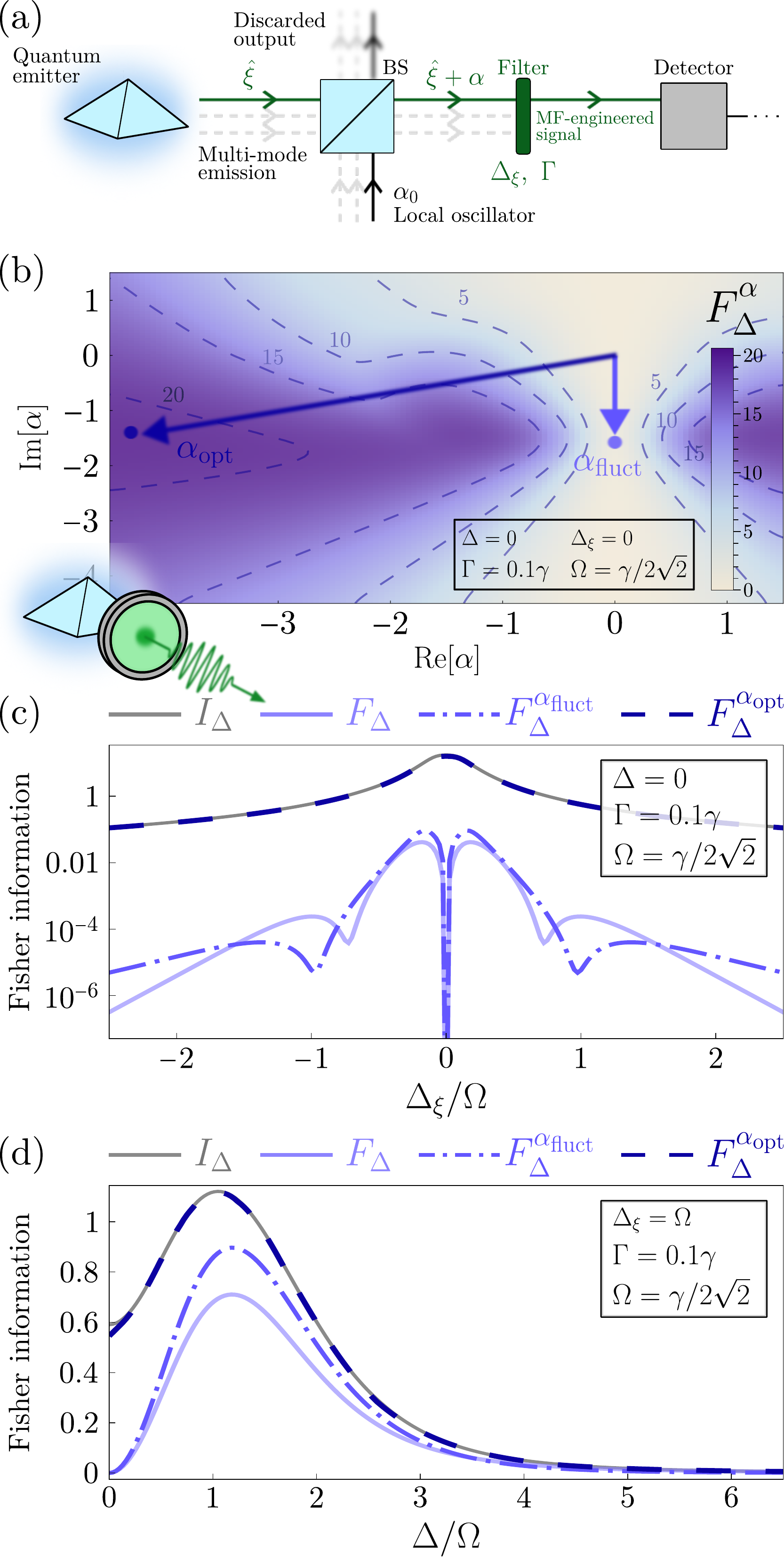}
		\caption{
			Optimizing quantum parameter estimation via mean-field engineering for a single sensor.
			(a) Mean-field engineering scheme: the multi-mode radiation is mixed with a coherent field via an unbalanced beam splitter in the limit of perfect transmittance, yielding an engineered signal that is frequency-filtered.
			(b) $F^\alpha_\Delta$ over the phase space of the local oscillator displacement.
			Two points are highlighted: 
			the displacement  $\alpha_\text{fluct}\approx-1.58 i$, and the optimal displacement $\alpha_{\text{opt}}\approx-4.3-1.4i$. 
			(c) $I_\Delta$ (grey solid), $F_\Delta$ (blue solid), $F_\Delta^{\alpha_{\text{fluct}}}$ (blue dot-dashed), and $F_\Delta^{\alpha_{\text{opt}}}$ (blue dashed) in the frequency domain ($\Delta_\xi$).
			(d) Same quantities as in (c), now plotted against the emitter-laser detuning ($\Delta$), with $\Delta_\xi=\Omega$.
			Parameters: (b-d) $\Omega=\gamma/2\sqrt{\gamma}$, $\Gamma=10^{-1}\gamma$, $\varepsilon=0.5$;  (b,c) $\Delta=10^{-6}\gamma$; (b) $\Delta_\xi=0$; (d) $\Delta_\xi=\Omega$. 
		}
		\label{fig:MF-engineering}
	\end{figure}

	Following the scheme presented in Refs.~\cite{KimUnlockingMultiphoton2025,BrachtTunableMultiphoton2025}, we implement theoretically this strategy by introducing a highly transmissive beam splitter (BS) prior to frequency-filtering and detection in which the target frequency mode, $\hat \xi$, is mixed with a local oscillator of amplitude $\alpha_0\in \mathbb{C}$ [see Fig.~\ref{fig:MF-engineering}(a)]. 
	In the limit of perfect transmittance ($T\rightarrow1$ and $R\rightarrow0$), the target frequency mode in one of the outputs is approximately given by
	$\hat \xi'\approx \hat \xi 
	+\alpha$,
	where $\alpha\equiv iR\alpha_0$. The other port, $\hat o\approx  \alpha_0$, is discarded.
	This scheme is formally equivalent to applying a coherent displacement operator, $\hat{D}(\alpha)=\exp(\alpha \hat \xi^\dagger-\alpha^*\hat \xi)$, on the sensor density matrix and is readily extended to the multi-sensor case (with the caveat of involving several coherent fields of different frequencies matching those of the filters). 
	The possible mode mismatch between the target frequency-filtered mode and the local oscillator~\cite{GuptaEffectImperfect2020,NotarnicolaQuantumCommunications2025} is incorporated into our analysis simply by treating $\alpha_0$ as the coherent fraction of the local oscillator that is mode-matched. There is no reduction in interference visibility, as modes other than the target mode at the BS output are removed by the frequency filters applied prior to detection.
	
	Figure~\ref{fig:MF-engineering}(b) shows the CFI as a function of the displacement $\alpha$ in phase space at a fixed frequency $\Delta_\xi$, evidencing that displacing the signal away from $\alpha=0$---which corresponds to standard photon-counting, illustrated by the blue solid curve in Fig.~\ref{fig:MF-engineering}(c)---can significantly enhance the CFI.
	Following the idea presented in Ref.~\cite{ZubizarretaCasalenguaTuningPhoton2020,KimUnlockingMultiphoton2025,BrachtTunableMultiphoton2025}, we first consider the displacement that completely cancels the coherent component and isolates quantum fluctuations by tuning the amplitude to (see Appendix~\ref{Appendix:MeanField}): 
	\begin{equation}
		\alpha_{\text{fluct}} =  \frac{\sqrt{\varepsilon \gamma \Gamma}\langle \hat{\sigma} \rangle}{ \Gamma/2 + i\Delta_\xi},
	\end{equation}
	shown as a point in Fig.~\ref{fig:MF-engineering}(b) and across the frequency domain in Fig.~\ref{fig:MF-engineering}(c), denoted by $F_\Delta^{\alpha_{\text{fluct}}}$ (dot-dashed). 
	Complete cancellation of the coherent component modifies the metrological performance, but it does not always result in an improvement; the effect depends on the spectral region: e.g., it improves sensitivity near resonance but not at $\Delta_\xi \approx \Omega$.
	More generally, we can optimize the measurement by choosing the displacement $\alpha_{\text{opt}}\equiv \arg [\max_{\alpha}  F_\theta^\alpha]$ that yields the maximum CFI, $F_\theta^{\alpha_{\text{opt}}}$. The optimum displacement is also indicated as a point  in Fig.~\ref{fig:MF-engineering}(b). 
	This protocol significantly enhances the CFI and can even saturate the quantum Fisher information across the full frequency range, $F_\theta^{\alpha_{\text{opt}}}\approx I_\theta$, as shown in Fig.~\ref{fig:MF-engineering}(c).
	Intuitively, interfering the signal with a tunable local oscillator effectively reshapes how the parameter dependence is distributed among the normally ordered moments that enter the CFI [see Eq.~\eqref{eq:FisherInform_Corr_OneSensor}]. By adjusting the displacement, one can enhance the relative change of these moments with respect to the parameter $\theta$, thereby altering---and potentially increasing---the achievable Fisher information.
	We emphasize that the local oscillator does not introduce new information; rather, the displacement modifies how the existing parameter dependence of $\hat\rho_\theta$ and $\partial_\theta \hat\rho_\theta$ appears in the normally ordered moments that determine the CFI.
	%
	%
	For this reason, neither $\alpha=0$ nor $\alpha_{\mathrm{fluct}}$ correspond to privileged operating points: they are simply two particular choices within a continuous set of possible displacements. The optimal displacement $\alpha_{\mathrm{opt}}$ emerges from fully exploiting this phase-space freedom to maximize the Fisher information, as illustrated in Fig.~\ref{fig:MF-engineering}(b). 

	Further information on the metrological impact of mean-field engineering is shown in Fig.~\ref{fig:MF-engineering}(d), where we compare the QFI and the CFIs across values of the estimated parameter, $\theta=\Delta$, for the three relevant displacements mentioned here: $\alpha = 0$ (standard photon-counting), $\alpha_{\text{fluct}}$ (isolation of fluctuations), and $\alpha_{\text{opt}}$ (optimal measurement). 
	All curves show a global maximum near $\Delta \approx \Omega$, explicitly showing that the most sensitive point is not the resonant case, as it was already reported in Ref.~\cite{GammelmarkFisherInformation2014}.
	As in panel (c), the optimally displaced CFI provides the maximum achievable information that saturates the QFI, with only minor deviations near resonance ($\Delta = 0$), thereby reaching the ultimate metrological limit.

	These results establish mean-field engineering, combined with frequency-resolved photo-detection, as a powerful method to maximize parameter sensitivity.
	In fact, beyond the examples presented in Fig.~\ref{fig:MF-engineering},  we confirm the validity of this technique by optimizing the performance shown earlier in Fig.~\ref{fig:ImpactSensorLinewidth}(a), where the optimized CFI (blue dashed curve) saturates the QFI across the full range of $\Gamma$. As we will see later, this strategy naturally extends to the multi-mode case, where one optimizes a coherent displacement for each filtered mode.

	We also observe that the QFI does not inherit the intricate spectral structure visible in the CFI, but instead exhibits a smooth, broad shape centered at the drive resonance.
	Intuitively, the optimal measurement extracts information most efficiently near resonance, where the collected mode overlaps maximally with the emitted radiation carrying information about $\theta$. Moving away from resonance reduces this overlap and decreases the achievable precision. 
	From a theoretical perspective, displaced photon counting does not span the full POVM space of the radiation, though varying the coherent displacement already explores a large measurement manifold that might contain the optimal measurement. Therefore, optimizing the CFI over $\alpha$ effectively recombines contributions from all correlator orders in Eq.~\eqref{eq:FisherInform_Corr_OneSensor}, suppressing cancellations and yielding smoother variations in parameter space.

	\section{Harnessing photon-photon correlations in quantum parameter estimation}
	\label{sec:sec4}
	
	In this section, we explore how frequency-resolved photon-photon correlations can be harnessed to enhance quantum parameter estimation.
	To this end, we go beyond single-channel detection and consider joint photo-detection by using two frequency-resolved sensors, as described by Eq.~\eqref{eq:cascadedmaster2} and illustrated in Fig.~\ref{fig:Fig1_setup}(a). For the sake of simplicity, we focus on standard photon-counting without mean-field engineering ($\vec \alpha=\vec 0$) unless stated otherwise.
	Joint photodetection enables access to the full joint photon-counting distribution $p(n_1, n_2|\theta)$, which encodes marginal statistics at each detector---$p(n_1|\theta)=\sum_{n_2}p(n_1, n_2|\theta)$ and $p(n_2|\theta)=\sum_{n_1}p(n_1, n_2|\theta)$---and their cross-correlations.
	Our goal is to assess whether these correlations provide a genuine metrological advantage that may justify the use of coincidence detection (e.g., with a Hanbury-Brown Twiss setup) over independent measurements.
	In this section, we also focus on estimating the qubit-laser detuning, $\theta=\Delta$. Results for other parameters, including the driving strength $\Omega$ and the decay rate $\gamma$, are provided in Appendix~\ref{Appendix:Two_sensor_Other_parameters}.

	\begin{figure*}
		\centering
		\includegraphics[width=1.\linewidth]{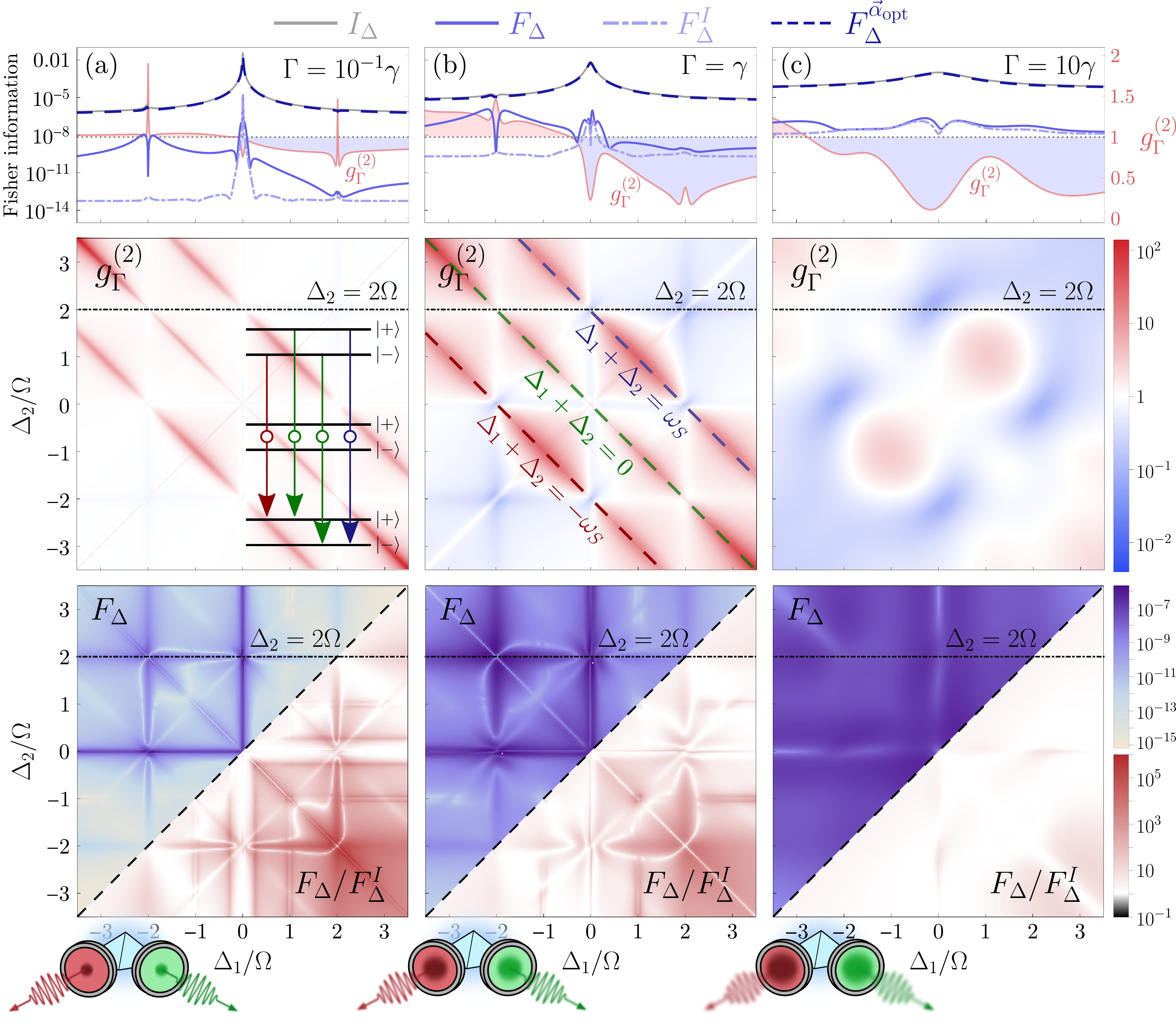}
		\caption{
			Frequency-resolved Fisher information for two sensors to estimate $\theta=\Delta$.
			Panels (a-c) correspond to different sensor linewidths $\Gamma=\{10^{-1},1 ,10\}\gamma$. Each column shares the same structure.
			Top row: $I_\Delta$ (grey solid), $F_\Delta$ (blue solid), $F_\Delta^I$ (blue dot-dashed), $F_\Delta^{\vec \alpha_{\text{opt}}}$ (blue dashed), and $g^{(2)}_\Gamma$ (red solid) in terms of one sensor-laser detuning $\Delta_1$, with the second fixed at  $\Delta_2=2\Omega$. 
			Middle row: $g^{(2)}_\Gamma$ in the frequency domain ($\Delta_1,\Delta_2$).
			Bottom row: $F_\Delta$ (upper-half) and $F_\Delta/F_\Delta^I$ (lower-half) in the frequency domain ($\Delta_1,\Delta_2$).
			Parameters: (a-c) $\Omega=10\gamma,\ \Delta=0,\ \varepsilon=0.5.$
		}
		\label{fig:Two-sensor-CFI}
	\end{figure*}

	The emission of a coherently driven two-level system, particularly in the Mollow regime, reveals genuine quantum correlations---evidence of non-classical light---that remain hidden in color-blind measurements but emerge through frequency-resolved detection~\cite{DelValleTheoryFrequencyFiltered2012,UlhaqCascadedSinglephoton2012,SanchezMunozViolationClassical2014,PeirisTwocolorPhoton2015,LopezCarrenoPhotonCorrelations2017,ZubizarretaCasalenguaConventionalUnconventional2020,LopezCarrenoEntanglementResonance2024,YangEntanglementPhotonic2025}.
	A central quantity that reflects such behaviour is the photon-photon correlations of the filtered emission by the zero-delay two-photon cross correlation function, which can be easily computed by means of the sensor method~\cite{DelValleTheoryFrequencyFiltered2012,CarrenoExcitationQuantum2016,CarrenoExcitationQuantum2016a}, 
	\begin{equation}
		g_\Gamma^{(2)}=\frac{\langle\hat \xi_1^\dagger \hat\xi_2^\dagger \hat \xi_1\hat \xi_2\rangle}{\langle \hat \xi_1^\dagger \hat \xi_1\rangle \langle \hat \xi_2^\dagger \hat \xi_2\rangle}.
	\end{equation}

\begin{figure}[t!]
		\centering
		\includegraphics[width=1.\linewidth]{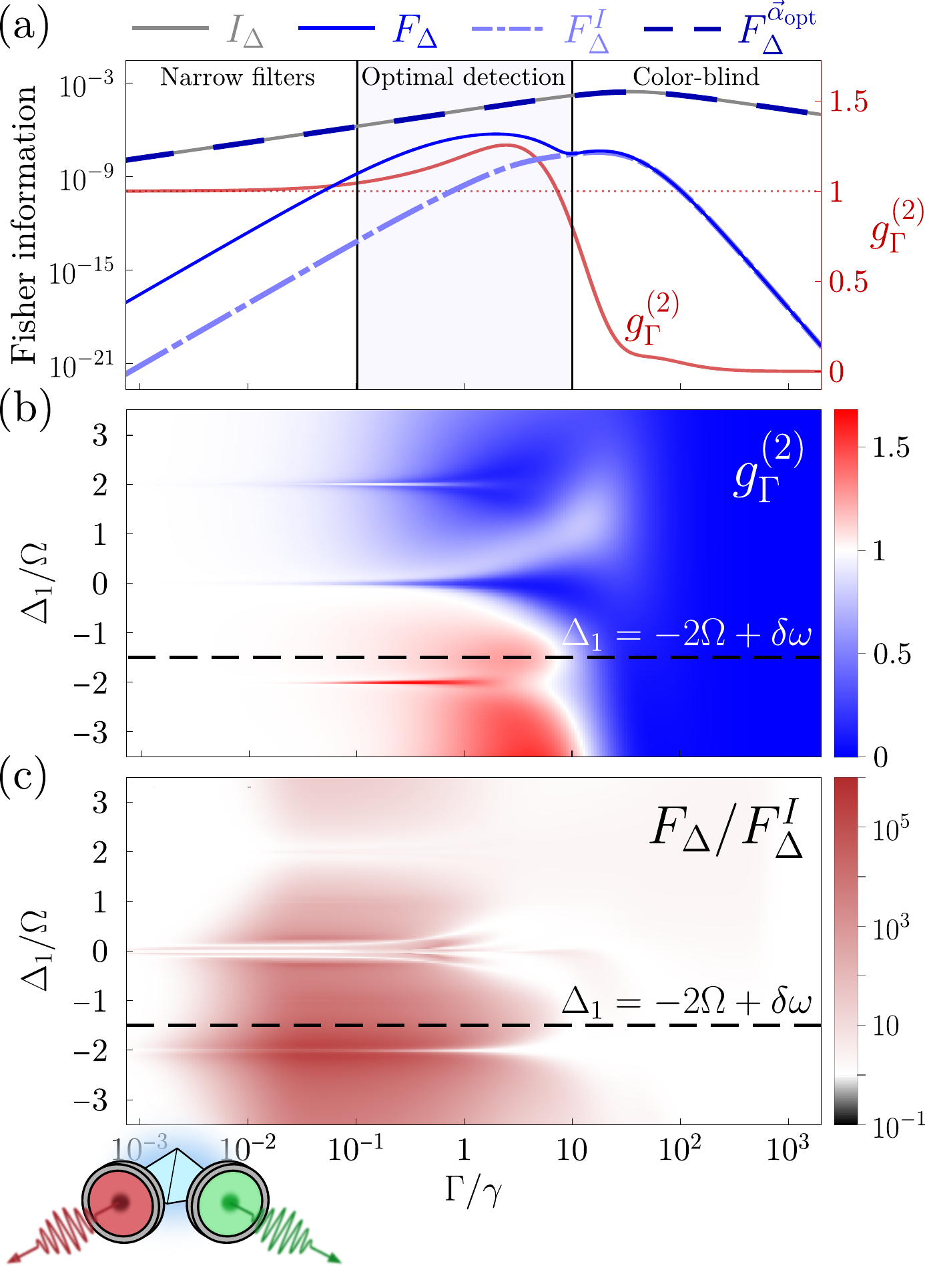}
		\caption{
			Impact of sensor linewidth $\Gamma$ on the performance of the estimation protocol for two sensors. 
			(a) $I_\Delta$ (grey solid), $F_\Delta$ (blue solid), $F_\Delta^I$ (blue dot-dashed), $F_\Delta^{\vec \alpha_{\text{opt}}}$ (blue dashed), and $g^{(2)}_\Gamma$ (red solid) in terms of $\Gamma$, for a fixed value of the sensors $(\Delta_1,\Delta_2)=(-2\Omega+\delta \omega,2\Omega)$. 
			The blue shaded region represents the optimal detection window.
			(b) $g^{(2)}_\Gamma$ and (c) $F_\Delta/F_\Delta^I$ in terms of the frequency of one sensor $\Delta_1$ and the sensor linewidth $\Gamma$, while having fixed $\Delta_2=2\Omega$. 
			%
			%
			Parameters: $\Delta=0,\ \Omega=10\gamma, \ \Delta_2=2\Omega, \ \delta\omega=0.5\Omega, \ \varepsilon=0.5$.
		}
		\label{fig:two-sensor-Gamma}
	\end{figure}

	This quantity provides evidences of multi-photon processes, such as the so-called leapfrog processes in the Mollow ladder~\cite{Gonzalez-TudelaTwophotonSpectra2013}, illustrated in the inset of the middle panel in Fig.~\ref{fig:Two-sensor-CFI}(a). %
	In the case of a two-photon leapfrog process, energy conservation sets the frequency of the two emitted photons, $\Delta_1$ and $\Delta_2$~\cite{Gonzalez-TudelaTwophotonSpectra2013,LopezCarrenoPhotonCorrelations2017}. They are determined by the conditions: (i)  $\Delta_1+\Delta_2=-\omega_S$ for the transition $|-\rangle \rightrightarrows |+\rangle$,  (ii) $\Delta_1+\Delta_2=0$ for the transition $|\pm\rangle \rightrightarrows |\pm\rangle$, and (iii) $\Delta_1+\Delta_2=\omega_S$ for the transition $|+\rangle \rightrightarrows |-\rangle$. 
	These two-photon transitions appear as antidiagonal lines in the insets of the second row in Fig.~\ref{fig:Two-sensor-CFI}, corresponding to superbunched emission (red regions). 
	The second row of Fig.~\ref{fig:Two-sensor-CFI} shows $g_\Gamma^{(2)}$ in frequency space $(\Delta_1, \Delta_2)$ for different sensor linewidths, $\Gamma = \{10^{-1}, 1, 10\} \gamma$. 
	As previously reported in Ref.~\cite{Gonzalez-TudelaTwophotonSpectra2013}:
	(i) narrow-band filters ($\Gamma \to 0$) tend to suppress correlations, yielding mostly uncorrelated emission (white);
	(ii) intermediate linewidths ($\Gamma\sim \gamma$) reveal  multi-photon structures, capturing rich quantum correlations;
	(iii) broadband filters ($\Gamma \to \infty$) blur spectral information---color-blind measurements---, recovering standard antibunching from a two-level system. These features are more clearly observed in Fig.~\ref{fig:two-sensor-Gamma}(a,b). 

	Similarly to Sec.~\ref{sec:sec3} [c.f. Fig.~\ref{fig:Single-sensor-Fisher-DrivingFrequency}(c)], the two-sensor CFI, $F_\Delta$, inherits a non-trivial spectral structure, shown in the upper-half of the maps shown in the third row in Fig.~\ref{fig:Two-sensor-CFI}.
	However, as was the case with the single-sensor CFI and the fluorescence spectrum, it does not directly mirror $g^{(2)}_\Gamma$ due to the inherently nonlinear dependence of $F_\theta$ on higher-order field moments.
	The joint probability distribution $p(n_1,n_2|\theta)$ is given by a sum of nonlinear functions of the normally ordered moments of both sensors (see Appendix~\ref{Appendix:ProbDistrbCorr}),
	\begin{equation}
		p(n_1,n_2|\theta)=\sum_{k_1\geq n_1}^{n_{\text{exc}}} \sum_{k_2\geq n_2}^{n_{\text{exc}}} c_{n_1,k_1} c_{n_2,k_2} \langle\hat \xi_1^{\dagger k_1} \hat \xi_2^{\dagger k_2} \hat \xi_1^{k_1} \hat \xi_2^{k_2} \rangle.
		\label{eq:joint_prob_corr}
	\end{equation}
	As a result, the generally displaced two-sensor CFI from Eq.~\eqref{eq:CFI_PC} can be explicitly written as
	\begin{widetext}
		\begin{equation}
			F_\theta^{\alpha_1,\alpha_2}=\sum_{n_1,n_2}^{n_{\text{exc}}} \frac{
				[\sum_{k_1\geq n_1}^{n_{\text{exc}}} \sum_{k_2\geq n_2}^{n_{\text{exc}}} c_{n_1,k_1} c_{n_2,k_2} \partial_\theta\langle\hat D^\dagger(\alpha_1,\alpha_2)\hat \xi_1^{\dagger k_1} \hat \xi_2^{\dagger k_2} \hat \xi_1^{k_1} \hat \xi_2^{k_2} D(\alpha_1,\alpha_2)\rangle]^2
			}{
				\sum_{k_1\geq n_1}^{n_{\text{exc}}} \sum_{k_2\geq n_2}^{n_{\text{exc}}} c_{n_1,k_1} c_{n_2,k_2} \langle\hat D^\dagger(\alpha_1,\alpha_2)\hat \xi_1^{\dagger k_1} \hat \xi_2^{\dagger k_2} \hat \xi_1^{k_1} \hat \xi_2^{k_2} D(\alpha_1,\alpha_2) \rangle
			},
			\label{eq:two-sensor-CFI-corr}
		\end{equation}
	\end{widetext}
	where we have defined $\hat D(\alpha_1,\alpha_2)\equiv \hat D(\alpha_1) \hat D(\alpha_2)$, with $\hat D(\alpha_i) =\exp[(\alpha_i\hat \xi^\dagger_i-\alpha_i^*\hat \xi_i)]$, since now each sensor can be independently displaced.
	While a direct mapping between $g^{(2)}_\Gamma$ and $F_\theta^{\alpha_1,\alpha_2}$ is not possible, the two-photon correlation function still offers physical insight into the structure of quantum correlations relevant for metrology.

	In previous works, regions of strong bunching---specially near leapfrog transitions---were linked to nonclassicality~\cite{SanchezMunozViolationClassical2014}, and generation of entanglement~\cite{LopezCarrenoEntanglementResonance2024,YangEntanglementPhotonic2025}. 
	However, $F_\theta$ responds to a more complex combination of multi-mode moments, making its structure more intricate.
	Here, instead, we find that, for the particular case of $\theta=\Delta$, the two-sensor CFI takes maximum values along the axes $(0,\Delta_2)$ and $(\Delta_1,0)$, and near leapfrog transitions, such as $(\Delta_1, \Delta_2) \approx (-2\Omega + \delta \omega, 2\Omega)$---where $\delta \omega$ is a small frequency variation---. This behavior is plotted by the blue curves in the top panels of Fig.~\ref{fig:Two-sensor-CFI}, having fixed $\Delta_2=2\Omega$.
	The first type of transition involves cascaded emission in which the two photons are emitted subsequently,  e.g., $(\Delta_1,\Delta_2)= (0,0)$, resulting in $|\pm\rangle\rightarrow |\pm\rangle \rightarrow|\pm \rangle$, and thus not involving virtual processes. 
	On the other hand, the second type reflects a quantum-enhanced process as it is due to a leapfrog transition.

	This statement can be clearly proven by comparing the two-sensor CFI with the case where cross-correlations are inaccessible and the joint probability distribution factorizes---e.g., when detection at both frequencies is performed independently---yielding an uncorrelated two-sensor CFI given by the sum of the individual CFIs from each detector,
	\begin{equation}
		F_\theta^{I}\equiv F_\theta[p(n_1|\theta)]+F_\theta[p(n_2|\theta)].
		\label{eq:uncorrelatedBound}
	\end{equation}
	Physically, this situation may occur when, for instance, a path-length mismatch in a the HBT setup [see Fig.~\ref{fig:Fig1_setup}(a)] introduces time delays large enough to erase any information about photon-photon correlations.

	In Fig.~\ref{fig:Two-sensor-CFI}, the top panels show $F_\Delta^I$ (blue dot-dashed) as a function of $\Delta_1$ for a fixed $\Delta_2 = 2\Omega$, while the lower-half panels in the third row display the ratio $F_\theta / F_\theta^I$, which quantifies the metrological gain from retaining cross-correlations compared to independent measurements.
	When $F_\theta / F_\theta^I>1$ (red regions), we have a metrological advantage, reaching up to five orders of magnitude near leapfrog resonances, as it is shown in the top panels via the blue dot-dashed curves ($F_\Delta^I$). 
	For instance, in the aforementioned region, $(\Delta_1,\Delta_2)\approx (-2\Omega+\delta \omega,+2\Omega)$, we find $F_\theta \approx 10^5 F_\theta^I$. 
	Generally, retaining cross-correlations ($F_\Delta$) outperforms the independent measurements ($F_\Delta^I$), as long as spectral resolution is guaranteed. 
	In the case of broadband filtering, the spectral information is washed out, resulting in $F_\Delta \approx F_\Delta^I$ (white regions).

	These features are further illustrated in Fig.~\ref{fig:two-sensor-Gamma}. In panel (a), both $F_\Delta$ (blue solid) and $F_\Delta^I$ (blue dot-dashed) are plotted as a function of $\Gamma$ at a representative leapfrog point $(\Delta_1, \Delta_2) = (-2\Omega + \delta\omega, 2\Omega)$, while panel (c) displays the ratio $F_\Delta/F_\Delta^I$ across $\Delta_1$ and $\Gamma$, clearly highlighting the spectral window where quantum correlations yield maximal metrological benefit.

	It is important to emphasize that although $F_\theta > F_\theta^I$ holds for the cases analyzed here, this is not generally guaranteed. One might intuitively expect that the joint probability distribution always offers a metrological advantage over uncorrelated measurements, as it captures more complete information. While this is indeed observed in Fig.~\ref{fig:Two-sensor-CFI} and in Fig.~\ref{fig:two-sensor-Gamma}, it does not hold generally.
	We have derived a general bound (see Appendix~\ref{Appendix:UncorrelatedBound}) establishing that:
	\begin{equation}
		F_\theta[p(n_1,n_2|\theta)]\geq \frac{1}{2} (F_\theta[p(n_1|\theta)]+F_\theta[p(n_2|\theta)]).
	\end{equation}
	In other words, one cannot assert that knowledge of the full joint probability distribution always guarantees a metrological advantage over performing
	uncorrelated measurements. While this result might seem counter-intuitive, it can be understood by considering the limiting case of two perfectly correlated (or anticorrelated) random variables: once one variable is known the other is straightforwardly inferred and carries no extra information. Therefore, in such cases, $p(n_1,n_2|\theta)$ does not provide more information than the product $p(n_1|\theta)p(n_2|\theta)$, which describes
	uncorrelated events each carrying a piece of information.
	
	Finally, we note that the mean-field engineering method---introduced earlier in Fig.~\ref{fig:ImpactSensorLinewidth}(a) and in Fig.~\ref{fig:MF-engineering}---extends naturally to the two-mode scenario. In this case, we must find the optimal displacement for each mode, $F_\theta^{\vec \alpha_\text{opt}}$, where $\vec\alpha_{\mathrm{opt}}=(\alpha_1^{\mathrm{opt}},\alpha_2^{\mathrm{opt}})$ denotes the optimal set of displacements.
	This definition generalizes straightforwardly to the N-sensor case, where the optimization is carried out over the full set of $N$ coherent displacements.
	The resulting optimized two-mode CFI is shown as blue dashed curves in the upper panels of Fig.~\ref{fig:Two-sensor-CFI} and in Fig.~\ref{fig:two-sensor-Gamma}(a). In both cases, $F_\theta^{\vec\alpha_{\mathrm{opt}}}$ exactly saturates the QFI, proving that mean-field engineering provides a general and effective strategy for optimizing the sensitivity of parameter estimation from displaced photon-counting measurements.

	\section{Generality of frequency-resolved quantum metrology}
	\label{sec:Generality}

	Throughout the manuscript, we focused on the simplest nontrivial quantum-optical system---a coherently driven two-level system---to clearly show the underlying mechanisms and the rich phenomenology enabled by multi-mode frequency filtering and mean-field engineering for quantum metrology.
	However, the framework we developed is fully general.
	%

	To illustrate this broad scope, we now briefly show how the same techniques apply to other relevant physical settings. As an example, we consider the estimation of a nonlinearity $\chi$: first in a single anharmonic oscillator---such as an optical cavity with Kerr nonlinearity or a transmon qubit---and then in a configuration with richer spectral features, where the same system is optomechanically coupled to a resonator yielding Raman sidebands. 
	\subsection{Example 1: Nonlinear harmonic oscillator (transmon qubit)}

\begin{figure}[b!]
		\centering
		\includegraphics[width=1.\linewidth]{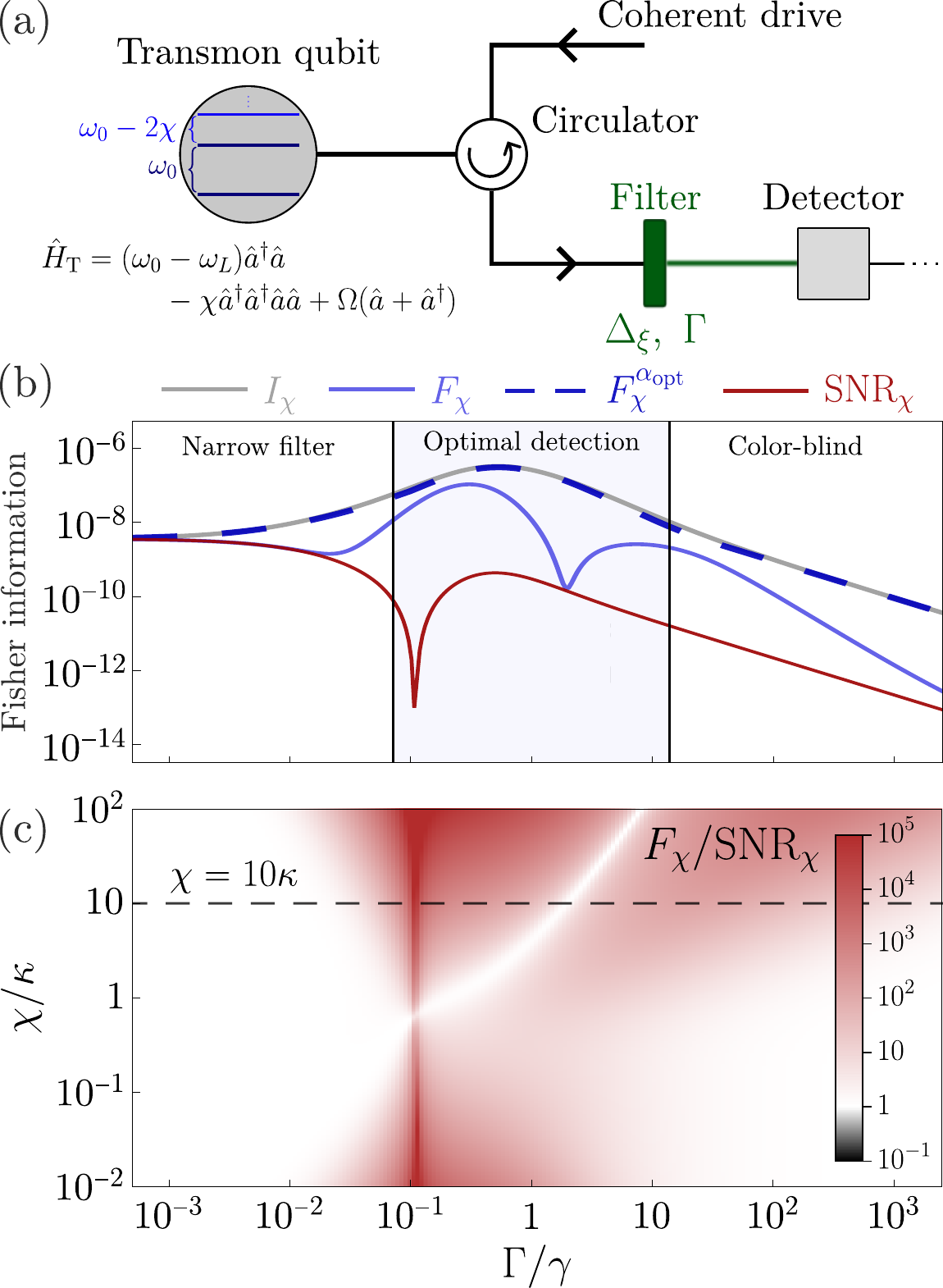}
		\caption{
			Frequency-resolved Fisher information for one-sensor for the estimation of $\theta=\chi$.
			(a) Sketch of the model:
			a transmon qubit is coupled to a waveguide in a reflection setup using a circulator. The coherent drive enters through the waveguide and excites the transmon, while the outgoing field is redirected toward the measurement port. The emitted radiation is subsequently frequency-filtered.
			(b, c) Impact of the sensor linewidth $\Gamma$ on the performance of the estimation protocol for a single sensor.
			(b) $I_\chi$ (grey solid), $F_\chi$ (blue solid),  $F_\chi^{\text{opt}}$ (blue dashed), and  $\text{SNR}_\chi$ (red solid) as a function of $\Gamma$, shown for a fixed non-linearity $\chi = 10 \kappa$.
			(c) $F_\chi/\text{SNR}_\chi$ as a function of $\chi$ and $\Gamma$.
			Parameters: $\Delta_a=0$, $\Omega=0.1\kappa$, $\varepsilon=1$, $\Delta_\xi=2\Omega$.
		}
		\label{fig:Example_TransmonQubit}
	\end{figure}

	In this first example, we consider a nonlinear harmonic oscillator (with annihilation operator $\hat a$) characterized by a resonant frequency $\omega_0$ and a Kerr-type anharmonicity $\chi$, that is coherently driven by a field with amplitude $\Omega$ and frequency $\omega_L$.  
	A specific physical setup described by this model is a superconducting transmon qubit coupled to a one-dimensional waveguide~\cite{YangEntanglementPhotonic2025,GuMicrowavePhotonics2017,GarciaRipollQuantumInformation2022} [see Fig.~\ref{fig:Example_TransmonQubit}(a)].  
	A circulator routes the incoming coherent drive towards the transmon, while directing the outgoing radiation to the detection stage.
	In the rotating frame of the coherent drive, and under the rotating wave approximation, the Hamiltonian of the driven transmon reads
	\begin{equation}
		\hat H_{\mathrm{T}}= \Delta_a \hat a^\dagger \hat a -\chi \hat a^{\dagger}\hat a^{\dagger}\hat a \hat a+ \Omega(\hat a+ \hat a^\dagger),
		\label{eq:transmon_Hamiltonian}
	\end{equation}
	where $\Delta_a\equiv \omega_a-\omega_L$ is the transmon-laser detuning.
	The anharmonicity $\chi$ quantifies the deviation of the transmon from an ideal harmonic oscillator. In the limit $\chi\to 0$, the system behaves as a linear resonator, while for $\chi\to\infty$ the higher excited states become strongly detuned ($\Delta E_{n+1,n}=\omega_0-2\chi n$), such that the dynamics effectively reduces to that of a two-level system.
	Physically, $\chi$ is directly related to the charging energy of the transmon, $E_C$, through $\chi \simeq -E_C$~\cite{GuMicrowavePhotonics2017,GarciaRipollQuantumInformation2022}. 
	%
	%
	
	%
	%
	%
	In the Markovian regime, the dissipative dynamics of the system---described by its reduced density matrix $\hat \rho_{\mathrm{T}}$---, is governed by the following master equation
	\begin{equation}
		\frac{d \hat \rho_{\mathrm{T}} }{dt}= -i [\hat H_{\mathrm{T}}, \hat \rho_{\mathrm{T}}] + \frac{\kappa}{2} \mathcal{D}[\hat a] \hat \rho_{\mathrm{T}},
		\label{eq:MasterEq_Transmon}
	\end{equation}
	where $\kappa$ is the relaxation rate of transmon into the waveguide.
	
	To analyze the metrological properties of the emitted radiation, we extend Eq.~\eqref{eq:MasterEq_Transmon} to model the cascaded system of transmon and ancillary sensor. 
	The joint system is described by the density matrix $\hat \rho_{\mathrm{T,1}}$, and the corresponding master equation is given by 
	\begin{multline}
		\frac{d \hat \rho_{\mathrm{T,1}} }{dt}=-i [\hat H_{\mathrm{T}}+\Delta_\xi \hat \xi^\dagger \hat \xi, \hat \rho_{\mathrm{T,1}}] + \frac{\kappa}{2} \mathcal{D}[\hat a] \hat \rho_{\mathrm{T,1}}  
		\\
		+ \frac{\Gamma}{2} \mathcal{D}[\hat \xi] \hat \rho_{\mathrm{T,1}}  -
		\sqrt{\varepsilon \kappa \Gamma} \left\{ [\hat \xi^\dagger, \hat a  \hat \rho_{\mathrm{T,1}} ] + [\hat \rho_{\mathrm{T,1}} \hat a^\dagger, \hat \xi] \right\}.
	\end{multline}
	Figure~\ref{fig:Example_TransmonQubit}(b) shows the CFI (blue solid) as a function of the sensor linewidth $\Gamma$ for a particular choice of the nonlinearity, $\chi=10 \kappa$. 
	As in the driven-TLS case [see Fig.~\ref{fig:ImpactSensorLinewidth}], the CFI exhibits an optimal detection window, here located approximately in the range $\Gamma_{\mathrm{opt}}\approx(10^{-1}\kappa,\,10\,\kappa)$, whereas in both the narrowband and broadband limits the CFI becomes suppressed.

	In the narrowband regime $(\Gamma\lesssim 10^{-1}\kappa)$, the CFI collapses onto the SNR (red solid), $F_\chi\approx\mathrm{SNR}_\chi$, indicating that the estimation is dominated by mean values. 
	Remarkably---and in contrast to the driven-TLS case [see Fig.~\ref{fig:ImpactSensorLinewidth}(a)]---both quantities (CFI and SNR) saturate the QFI (grey solid). This proves that, in the narrowband sensor limit, single-detector measurements are already optimal for estimating the transmon nonlinearity $\chi$.
	In the opposite broadband limit ($\Gamma\gtrsim 10\kappa$), the CFI approaches the SNR as $\Gamma\to\infty$ (not shown), as expected for color-blind detection, where spectral resolution is lost.
	The QFI remains greater than the CFI for all $\Gamma$, including in the broadband regime, implying the existence of a POVM that outperforms direct photon-counting. 
	By applying mean-field engineering prior to detection [see Fig.~\ref{fig:MF-engineering}(a)], we obtain an optimized CFI that saturates the QFI across the full range of $\Gamma$, such that $I_\chi\simeq F_\chi^{\alpha_{\mathrm{opt}}}$ (blue dashed). This result provides an additional validation of the mean-field engineering approach, showing that even outside the optimal detection region, the estimation sensitivity can still be significantly enhanced.

	The above analysis focused on a specific value of the transmon nonlinearity.
	A similar picture emerges for other values of $\chi$, as shown in Fig.~\ref{fig:Example_TransmonQubit}(c), which displays the ratio $F_\chi/\mathrm{SNR}\chi$ as a function of $\chi$ and $\Gamma$.
	Regions where this ratio exceeds unity (in red) identify spectral windows in which exploiting the full photon-number distribution provides a metrological advantage over relying only on the mean value, while white regions mark regimes where the field is essentially coherent.
	%
	%
	Generally, the same behavior reported in panel (b) is present here: a narrowband region where $F\chi\approx\mathrm{SNR}\chi$, an optimal detection window where $F\chi\gg\mathrm{SNR}\chi$, and a broadband limit in which $F\chi\approx\mathrm{SNR}\chi\rightarrow0$.
	This shows that the metrological structure of the emitted radiation, when used to estimate the anharmonicity, exhibits a similar dependence on $\chi$ both in the nearly harmonic regime ($\chi \to 0$) and in the strongly anharmonic, qubit-like regime ($\chi \to \infty$). We emphasize, however, that although the ratio $F_\chi/\mathrm{SNR}_\chi$ varies smoothly with $\chi$, the absolute values of CFI and SNR can differ significantly across regimes. In addition, we observe a peculiar feature, located near $(\chi,\Gamma)\approx(1,10^{-1})\kappa$ and extending over larger $\Gamma$, where $F_\chi\approx\text{SNR}_\chi$.
	%
	%
	Further analysis of this system---and of the particular effect just mentioned---is left for a future work.

	\subsection{Example 2: Optomechanical system}
	
	A natural next step beyond the isolated nonlinear oscillator considered in the transmon example above is to explore systems where such nonlinearities interact with additional degrees of freedom.
	A paradigmatic and technologically relevant example is provided by cavity optomechanics (OM), where the appearance of Raman sidebands---central to many applications in molecular spectroscopy~\cite{OrlandoComprehensiveReview2021}---highlights the broad applicability of our framework in a platform exhibiting an even richer spectral structure.
	In this context, we note that spectral measurements in optomechanical platforms have also been explored from a complementary perspective, namely to establish metrological bounds for the estimation of time-varying signals and external forces~\cite{TsangFundamentalQuantum2011,NgSpectrumAnalysis2016}
	.

	In these platforms, the nonlinear coupling between mechanical resonator and optical mode produces a characteristic emission spectrum composed of Stokes and anti-Stokes sidebands associated with phonon creation and annihilation processes~\cite{AspelmeyerCavityOptomechanics2014}.
	Frequency-resolved correlations have been studied in these platforms in the case of a general OM system~\cite{SchmidtFrequencyresolvedPhoton2021}, or in specific molecular realizations~\cite{Martinez-GarciaCoherentElectronVibron2024,MoradiKalardePhotonAntibunching2025}, where a rich landscape of frequency-resolved correlations is unraveled due to the intrinsic non-linearities of the system. 

\begin{figure}[b!]
		\centering
		\includegraphics[width=1.\linewidth]{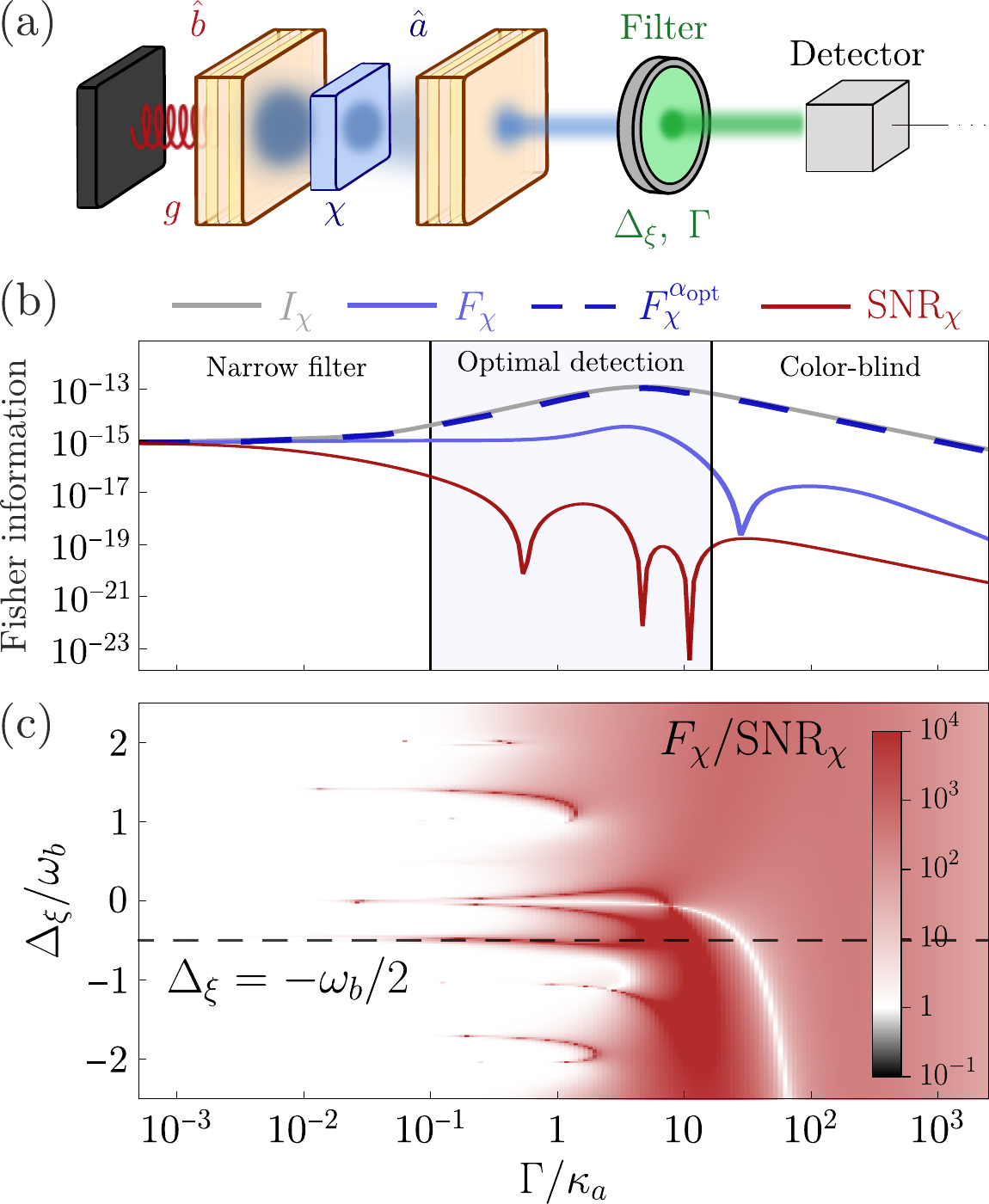}
		\caption{
			Frequency-resolved Fisher information for one-sensor for the estimation of $\theta=\chi$.
			(a) Sketch of the model: 
			a mechanical resonator ($\hat b$) coupled, via a nonlinear interaction of strength $g$, to an optical cavity mode ($\hat a$) that exhibits an intrinsic nonlinearity $\chi$. The cavity is also coherently driven (not shown in the sketch). The output field of the cavity is spectrally filtered at frequency $\Delta_\xi$ with resolution $\Gamma$.
			(b, c) Impact of the sensor linewidth $\Gamma$ on the performance of the estimation protocol for a single sensor.
			(b) $I_\chi$ (grey solid), $F_\chi$ (blue solid),  $F_\chi^{\text{opt}}$ (blue dashed), and  $\text{SNR}_\chi$ (red solid) as a function of $\Gamma$, shown for a fixed sensor frequency $\Delta\xi = -\omega_b/2$.
			(c) $F_\eta/\text{SNR}_\chi$ as a function of $\Delta\xi$ and $\Gamma$.
			Parameters: $\Delta_a=0$, $\Omega=0.1\kappa_a$, $\chi=10^2\kappa_a$, $\omega_b=5\kappa_a$, $\kappa_b=10^{-2}\kappa_a$, $g=\kappa_a$, $n_{\text{th}}=10^{-2}$, $\varepsilon=0.1$.
		}
		\label{fig:Example_DrivenOMresonator}
	\end{figure}

	We consider a OM system with a single-mode cavity [see Fig.~\ref{fig:Example_DrivenOMresonator}(a)], with annihilation operator $\hat a$, frequency $\omega_a$ and Kerr nonlinearity $\chi$, which is coherently driven with amplitude $\Omega$ at frequency $\omega_L$. 
	This optical mode is nonlinearly coupled to a mechanical resonator (with annihilation operator $\hat b$) of frequency $\omega_b$ via an optomechanical interaction of strength $g$.   
	In the rotating frame of the coherent drive, and under the rotating wave approximation, the Hamiltonian of the OM system takes the form
	\begin{multline}
		\hat H_{\mathrm{OM}}= \Delta_a \hat a^\dagger \hat a +\chi \hat a^{\dagger}\hat a^{\dagger}\hat a \hat a+ \Omega(\hat a+ \hat a^\dagger)\\
		+ \omega_b \hat b^\dagger \hat b+g \hat a^\dagger \hat a (\hat b + \hat b^\dagger),
		\label{eq:OM_Hamiltonian}
	\end{multline}
	where $\Delta_a\equiv \omega_a-\omega_L$ is the cavity-laser detuning.
	Dissipation originates from photon loss at rate $\kappa_a$, phonon loss at rate $\kappa_b$, and thermal phonon excitations characterized by $n_{\mathrm{th}}=(e^{\omega_b/k_BT}-1)^{-1}$, where $k_B$ is the Boltzmann constant and $T$ is the temperature of the phonon bath.
	In the Markovian regime, the dissipative dynamics of the OM system---described by its reduced density matrix $\hat \rho_{\mathrm{OM}}$---, is governed by the following master equation
	\begin{multline}
		\frac{d \hat \rho_{\mathrm{OM}} }{dt}= -i [\hat H_{\mathrm{OM}}, \hat \rho_{\mathrm{OM}}] + \frac{\kappa_a}{2} \mathcal{D}[\hat a] \hat \rho_{\mathrm{OM}} \\+\frac{(n_{\mathrm{th}}+1)\kappa_b}{2} \mathcal{D}[\hat b] \hat \rho_{\mathrm{OM}}
		+\frac{n_{\mathrm{th}}\kappa_b}{2} \mathcal{D}[\hat b^\dagger] \hat \rho_{\mathrm{OM}}.
		\label{eq:MasterEq_Optomechanics}
	\end{multline}
	As we did in the previous example, we extend Eq.~\eqref{eq:MasterEq_Optomechanics} by including the description of the ancillary sensor, such that the corresponding master equation of the joint system---described by the density matrix $\hat \rho_{\mathrm{OM,1}}$---is given by 
	\begin{equation}
		\frac{d \hat \rho_{\mathrm{OM,1}} }{dt}=(\mathcal{\hat L}_{\mathrm{OM}} +\mathcal{\hat L}_{\mathrm{1}}) \hat \rho_{\mathrm{OM,1}}.
	\end{equation}
	Here, $\mathcal{\hat L}_{\mathrm{OM}}$ is the Liouvillian identified from Eq.~\eqref{eq:MasterEq_Optomechanics}, and $\mathcal{\hat L}_{\mathrm{1}}$ is the Liouvillian describing the ancillary sensor and the cascaded coupling:
	\begin{multline}
		\mathcal{\hat L}_{\mathrm{1}}\hat \rho_{\mathrm{OM,1}}\equiv -i [\Delta_\xi \hat \xi^\dagger \hat \xi, \hat \rho_{\mathrm{OM,1}}]+ \frac{\Gamma}{2} \mathcal{D}[\hat \xi]\hat \rho_{\mathrm{OM,1}}
		\\
		-
		\sqrt{\varepsilon \kappa_a \Gamma} \left\{ [\hat \xi^\dagger, \hat a  \hat \rho_{\mathrm{OM,1}} ] + [\hat \rho_{\mathrm{OM,1}} \hat a^\dagger, \hat \xi] \right\}.
	\end{multline}

	Compared to the driven TLS and the transmon qubit, the optomechanical system contains a substantially larger parameter space,
	$\theta\in \{\Delta_a,\chi,\Omega,\omega_b,g,\kappa_a,\kappa_b,T\}$,
	reflecting the richer underlying dynamics.
	A full metrological characterization of all these parameters lies far beyond the scope of this section.
	For illustrative purposes---and in continuity with the previous example---we focus on the estimation of the optical nonlinearity, $\theta=\chi$.

	In close analogy with the driven-TLS and transmon examples, the optomechanical system also exhibits a well-defined optimal detection window in which the CFI is maximized, while both the narrowband and broadband limits force the CFI to collapse onto the SNR, as shown in Fig.~\ref{fig:Example_DrivenOMresonator}(b).
	However, the OM nonlinearities generate a richer spectral structure~\cite{SchmidtFrequencyresolvedPhoton2021}.
	This is revealed in  Fig.~\ref{fig:Example_DrivenOMresonator}(c), which displays the ratio $F_\chi/\mathrm{SNR}_\chi$ as a function of both $\Delta_\xi$ and $\Gamma$.
	As expected, the ratio tends to unity in the narrowband limit (white regions), and would approach unity again in the deep broadband regime (not shown).
	Importantly, the best metrologically informative frequencies do not coincide with the physical resonances of the source. In the OM system [see Eq.~\eqref{eq:OM_Hamiltonian}], Stokes and anti-Stokes resonances occur at integer multiples of the mechanical frequency, $\pm m\omega_b$, yet optimal regions for estimating $\chi$ appears at other frequencies, e.g., at $\Delta_\xi=-\omega_b/2$ [as shown in panel (b)].
	As it already occurred with the driven-TLS [see Fig.~\ref{fig:Single-sensor-Fisher-DrivingFrequency}(a,c)], this counter-intuitive behavior stems from the fact that the CFI depends on a nontrivial interplay of derivatives of all normally ordered correlators of the filtered field [see Eq.~\eqref{eq:FisherInform_Corr_OneSensor}].

	Thus, although the scaling of the CFI with the sensor linewidth $\Gamma$ mirrors the behavior observed in the previous cases, the optomechanical setup shows how complex spectral structures---originating from mechanical nonlinearities---can be exploited to enhance parameter estimation.

	Our use of a single-sensor configuration and a single parameter in these examples was chosen to provide a simple, self-contained demonstration of how our framework applies to systems that are fundamentally different from the driven TLS on which most of our analysis was focused.
	A full exploration of the multi-parameter landscape and multi-sensor scenarios in such platforms is beyond the  scope of this text. Nevertheless, the results shown here already highlight the potential interest of exploring the metrological potential of frequency-filtered modes in complex quantum-optical systems.

	\section{Conclusions}
	\label{sec:sec5}
	We have presented a quantum-metrology framework that harnesses the spectral structure of radiation continuously emitted by open quantum systems. Working with the physical description of the quantum state of the frequency-filtered mode, we used the Fisher information to quantify the 
	precision gained from exploiting the full photon-counting distribution of frequency-filtered modes, as well as the correlations between distinct frequency components revealed through coincidence measurements. Extending this analysis across a range sensor frequencies and linewidths, we identified an optimal filtering regime---where neither spectral nor temporal information is lost---that maximizes sensitivity.

	Additionally, we analyzed the metrological potential of a mean-field engineering strategy that coherently displaces the detected signal. We showed that, by appropriately tuning the displacement, the CFI can reach the quantum Fisher information (QFI) of the sensors, effectively realizing the optimal POVM for the estimation task.
	We further extended our analysis to joint photo-detection via two frequency-resolved sensors. Here, we showed that photon-photon correlations can significantly enhance estimation precision. 
	
	Our results establish a physically grounded and versatile platform for quantum parameter estimation based on photon-counting and mean-field engineering of frequency-filtered radiation. This framework opens new possibilities in quantum sensing, quantum spectroscopy, and photonic quantum technologies.

	Several promising directions emerge from this work. One compelling avenue is the extension of our framework to more complex quantum systems, such as strongly nonlinear cavity-QED systems or systems operating near criticality, where richer dynamical behavior may unlock enhanced metrological performance. Another natural question is to what extent higher-order correlations—beyond second-order photon statistics—can further boost the information extractable from quantum light. Together, these directions open a broad and fertile landscape for advancing quantum metrology at the intersection of theory and experiment.

	\acknowledgments
	The authors are thankful to Alejandro González Tudela, Roberto Di Candia, Simone Felicetti, Maryam  Khanahmadi and Victor Rueskov Christiansen for insightful discussions. C. S. M. acknowledges support by the project PID2023-149969NA-100 funded by the Spanish Agencia Estatal de Investigación MICIU/AEI/10.13039/501100011033, and by a 2025 Leonardo Grant for Scientific Research and Cultural Creation from the BBVA Foundation.
	A.V.V. also acknowledges support from the Swedish Foundation for Strategic Research (grant number FFL21-0279).
	
	\appendix

	\section{Derivation of the one and two-sensors cascaded master equations}
	\label{Appendix:Derivation-MasterEq}
	
	In this section, we derive step-by-step the one- and two-sensors cascaded master equation presented in Eq.~\eqref{eq:cascadedmaster1} and Eq.~\eqref{eq:cascadedmaster2}, respectively. %
	To do this, we introduce general derivations for the cascaded systems, and later we identify each term with the corresponding elements from our setup.

	\subsection{Single-sensor cascaded master equation}
	
	In a cascaded quantum system, the output field of a system (the source) is fed into the input of another
	system (the target), without any back-action from the target to the source. This enables the description of setups driven by general sources of light.
	This is formally described by the cascaded formalism~\cite{GardinerDrivingQuantum1993,GardinerDrivingAtoms1994,CarmichaelQuantumTrajectory1993,GardinerQuantumNoise2004}.
	
\begin{figure}[t!]
		\centering
		\includegraphics[width=0.88\linewidth]{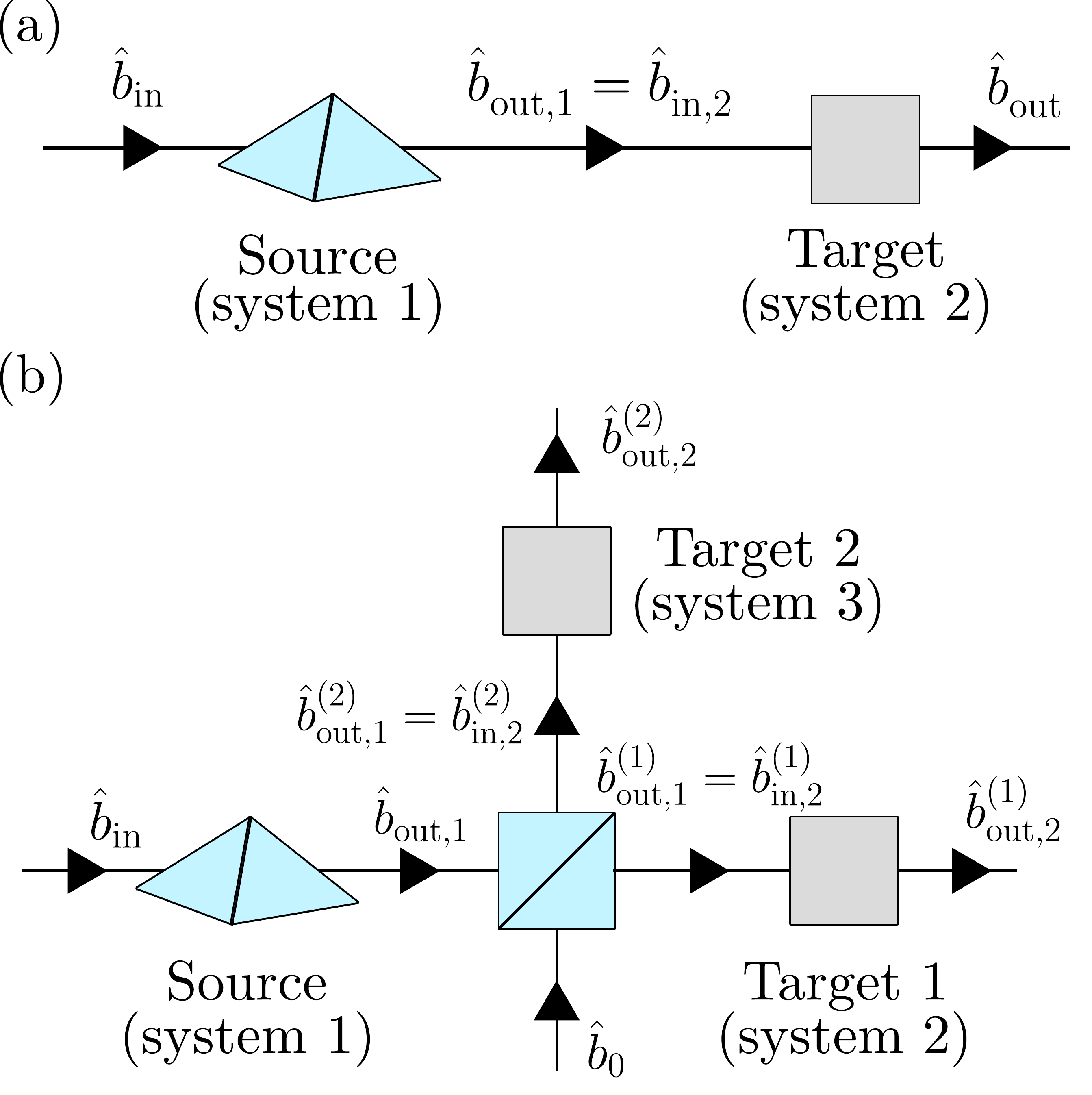}
		\caption{
			Schematic representation of cascaded systems.
			(a) A source (System 1), driven by an input field $\hat b_{\text{in}}$, feeds a target (System 2) with its own output field, such that $\hat b_{\text{out},1} = \hat b_{\text{in},2}$. This forms the simplest cascaded system, with a final output field $\hat b_{\text{out}}$.
			(b) A source (System 1), also driven by an input field $\hat b_{\text{in}}$, feeds two targets (System 2 and 3) by splitting its output field using a balanced beam splitter:  $\hat b^{(1)}_{\text{out},1}=\hat b^{(1)}_{\text{in},2}$ and $\hat b^{(2)}_{\text{out},1}=\hat b^{(2)}_{\text{in},2}$, respectively.
			Consequently, the full system features two output fields from the targets: $\hat b^{(1)}_{\text{out},2}$ and $\hat b^{(2)}_{\text{out},2}$.
		}
		\label{fig:CascadedSetup}
	\end{figure}

	The description of a cascaded quantum system is easily understood in terms of quantum Langevin equations, where a driving field $\hat b_{\text{in},1}(t)$ drives the source, giving rise to an output field $\hat b_{\text{out},1}(t)$, which, after a propagation
	time $\tau$, becomes the input field $\hat b_{\text{in},2}(t)$ of the second system.
	In this situation, let $\hat a_1$, $\hat a_2$ be operators for the two systems, and let $\hat H=\hat H_1+\hat H_2$ be
	the sum of their independent free Hamiltonians. In this configuration, the
	quantum Langevin equation for $\hat a_i$ (with $i=1,2$) reads:
	\begin{multline}
		\frac{d \hat a_i}{d t} = -i[\hat a_i, \hat H_i]-[\hat a_i, \hat c^\dagger_i] \left( \frac{\Gamma_i}{2} \hat c_i + \sqrt{\Gamma_i}\hat b_{\text{in},i} \right) 
		\\ +\left( \frac{\Gamma_i}{2} \hat c^\dagger_i + \sqrt{\Gamma_i}\hat b^\dagger_{\text{in},i} \right)[\hat a_i, \hat c_i].
	\end{multline}
	The cascaded coupling implies that the output field from the first system, $\hat b_{\text{out},1}=\hat b_{\text{in},1}+\sqrt{\Gamma_1}\hat c_1$, feds the second system after a time $\tau$.  We can express this cascaded effect by writing the input field of system $2$ as
	\begin{equation}
		\hat b_{\text{in},2}=\hat b_{\text{out},1}(t-\tau)=\hat b_{\text{in},1}(t-\tau)+\sqrt{\Gamma_1}\hat c_1(t-\tau),
		\label{eq:CascadedInputRelation}
	\end{equation}
	so that the output of the combined system reads:
	\begin{multline}
		\hat b_{\text{out},2}=\hat b_{\text{in},2}(t-\tau)+\sqrt{\Gamma_2}\hat c_2(t-\tau)
		\\
		=b_{\text{in},1}(t-\tau) +\sqrt{\Gamma_2}\hat c_2(t-\tau)+\sqrt{\Gamma_1}\hat c_1(t-\tau).
		\label{eq:CascadedOutputRelation}
	\end{multline}
	We now have operators evaluated at two different times, $t$ and $t-\tau$.  However, since we are considering a one-way driving scenario and $\tau > 0$, the effect of this retarded time is merely a time shift of the origin of the time axis for the second system. Thus, we can set $\tau \rightarrow 0^+$ without losing generality while maintaining causality~\cite{GardinerQuantumNoise2004}. We also note that, given the cascaded input-output relations in~\cref{eq:CascadedInputRelation,eq:CascadedOutputRelation}, we can identify the input and output fields of the combined systems just as $\hat b_{\text{in}}\equiv \hat b_{\text{in},1}$ and $\hat b_{\text{out}}\equiv \hat b_{\text{out},2}$ [see Fig.~\ref{fig:CascadedSetup}(a)].
	Finally, by considering a general operator $\hat a$, which may act either in the Hilbert space of the source or the target, we obtain the cascaded quantum Langevin equation:
	\begin{widetext}
		\begin{equation}
			\frac{d \hat a}{d t} = -i[\hat a, \hat H] - \sqrt{\Gamma_1 \Gamma_2} \left( [\hat a, \hat c^\dagger_2]\hat c_1- \hat c_1^\dagger[\hat a, c_2] \right)
			+\sum_{i=1}^2 \left[ 
			-[\hat a, \hat c_i^\dagger]\left( \frac{\Gamma_i}{2} \hat c_i + \sqrt{\Gamma_i}\hat b_{\text{in}} \right)
			+
			\left( \frac{\Gamma_i}{2} \hat c^\dagger_i + \sqrt{\Gamma_i}\hat b^\dagger_{\text{in}} \right)[\hat a, \hat c_i]
			\right].
			\label{eq:CascadedLangevin}
		\end{equation}%
	\end{widetext}

	Although the cascaded quantum Langevin equation gives an elegant description of the physics involved, it is sometimes more practical to convert this into the appropriate Quantum Ito equation~\cite{GardinerQuantumNoise2004,GardinerStochasticMethods2009}, and then obtain the corresponding master equation. 
	To achieve this, we must treat the input operator as a noise operator by replacing
	$
	\hat b_{\text{in}}(t)dt \rightarrow d\hat B(t),
	$ 
	where $ d\hat B(t)$ is the quantum Wiener increment, fulfilling the quantum Ito algebra~\cite{GardinerQuantumNoise2004,GardinerStochasticMethods2009,WisemanQuantumMeasurement2014}:
	\begin{subequations}
		\begin{align}
			&d\hat B(t)  d\hat B^\dagger(t)=dt, \\
			& d\hat B^\dagger(t) d\hat B(t)=0, \\
			& d\hat B(t)d\hat B(t)=d\hat B^\dagger(t)d\hat B^\dagger(t)=0.
		\end{align}
		\label{eq:ItoAlgebra}
	\end{subequations}
	All other products, including $dt d\hat B(t)$ or $dt d\hat B^\dagger(t)$, and higher orders are set to zero. We note that these quantum Ito rules are valid when one takes the expected value of the operators. Hence, using the quantum Ito algebra and taking the expected value of the quantum Langevin equation in Eq.~\eqref{eq:CascadedLangevin}, we obtain the cascaded master equation:
	\begin{multline}
		\frac{d \rho}{d t}=-i [\hat H,\hat \rho] + \sum_i^2 \frac{\Gamma_i}{2}\mathcal{D}[\hat c_i]\hat \rho
		\\
		- \sqrt{\Gamma_1 \Gamma_2}\left( [\hat c_2^\dagger , \hat c_1 \hat \rho]+ [\hat \rho \hat c_1^\dagger, \hat c_2] \right).
		\label{eq:CascadedMaster}
	\end{multline}
	This equation contains the standard terms of a Lindblad master equation:  a term for the free evolution governed by the Hamiltonian $\hat H$, and Lindblad terms for both the source and the target, with corresponding dissipative rates $\Gamma_1$ and $\Gamma_2$, respectively. Additionally, it features an extra term that accounts for the asymmetric cascaded coupling between the source and the target.
	We can see from Eq.~\eqref{eq:CascadedMaster} that it is exactly the same as Eq.~\eqref{eq:cascadedmaster1} by identifying:
	$\hat H \rightarrow \hat H + \Delta_\xi \hat \xi^\dagger \hat \xi$, $\hat c_1\rightarrow \hat \sigma$,  $\hat c_2\rightarrow \hat \xi$, $\Gamma_1 \rightarrow \gamma$,   $\Gamma_2\rightarrow \Gamma$, and $\sqrt{\Gamma_1\Gamma_2} \rightarrow \sqrt{\varepsilon \gamma\Gamma}$.

	\subsection{Two-sensor cascaded master equation}

	As discussed in the main text, extending the cascaded formalism to multiple target systems requires careful consideration, as different configurations lead to both quantitatively and qualitatively different dynamics. 
	%
	%
	In this work, we focus on a specific configuration in which the emission from the source (system 1) is split via a balanced 50:50 beam splitter, such that it simultaneously drives two target systems (systems 2 and 3), as illustrated in Fig.~\ref{fig:Fig1_setup}(a) and Fig.~\ref{fig:CascadedSetup}(b).
	This setup ensures that the sensors (targets) remain physically independent, allowing us to treat them as distinct subsystems. Accordingly, the Hilbert space factorizes as $\mathcal{H}=\mathcal{H}_1\otimes\prod_{\alpha=1}^2\mathcal{H}^{(\alpha)}_2$, such that $\hat c_1$ is an operator of the source, and $\hat c_2^{(\alpha)}$ stands as operators of the two independent sensors, with $\alpha = 1, 2$ labeling the respective targets.

	The insertion of the 50:50 beam splitter is a key distinction from previous derivations~\cite{YangEntanglementPhotonic2025}. This setup introduces vacuum contributions, denoted by $\hat b_0(t)$, into the output fields of the beam-splitter, now labeled as $\hat b_{\text{out},1}^{(\alpha)}$ for $\alpha=1,2$.
	Consequently, the beam splitter alters the structure of the cascaded interaction terms. By identifying the input fields to the targets as $\hat b_{\text{in},2}^{(\alpha)} = \hat b_{\text{out},1}^{(\alpha)}$, the inputs to the two sensors take the form:
	\begin{equation}
		\begin{pmatrix}
			\hat b_{\text{in},2}^{(1)} \\		
			\hat b_{\text{in},2}^{(2)}
		\end{pmatrix}
		=
		\frac{1}{\sqrt{2}}	\begin{pmatrix}
			1 & 1 \\
			1 & -1
		\end{pmatrix}
		\begin{pmatrix}
			\hat b_{\text{out},1} \\		
			\hat b_0
		\end{pmatrix}.
	\end{equation}
	Recalling that the output of the source is given by $\hat b_{\text{out},1} = \sqrt{\Gamma_1} \hat c_1 + \hat b_{\text{in}}$, the input fields to the two targets become:
	\begin{equation}
		\hat b_{\text{in},2}^{(\alpha)}(t)  \equiv \frac{1}{\sqrt{2}}[\sqrt{\Gamma_1}\hat c_1(t)+ \hat	b_{\text{in}}(t)+\hat b_0^{(\alpha)}(t)],
	\end{equation}
	where we have defined $\hat b_0^{(1/2)}(t)\equiv\pm \hat  b_0(t)$ to unify the notation.
	Since the output of the first system is equally split between the two targets, we still obtain three quantum Langevin equations, corresponding to the operators $\{\dot{\hat a}_1, \dot{\hat a}^{(1)}_2, \dot{\hat a}^{(2)}_2\}$, such that each target evolves independently, receiving a distinct vacuum contribution from the beam splitter.  Particularly, the dynamics of each target differ only in the specific vacuum input $\hat b_0^{(\alpha)}$, with $\alpha = 1, 2$, introduced by the beam splitter:
	\begin{widetext}
		\begin{subequations}
			\begin{align}
				\dot {\hat a}_1=&-i[\hat a_1,\hat H]  -[\hat a_1,\hat c_1^\dagger]  \left( \frac{1}{2}\Gamma_1 c_1 + \sqrt{\Gamma_1}\hat b_{\text{in}}   \right)
				+ \left( \frac{1}{2}\Gamma_1 \hat c_1^\dagger + \sqrt{\Gamma_1}\hat b^\dagger_{\text{in}}   \right) [\hat a_1,\hat c_1], \\
				\dot {\hat a}^{(\alpha)}_2=&-i[\hat a^{(\alpha)}_2,\hat H] -[\hat a^{(\alpha)}_2,{\hat c^{(\alpha)\dagger}_2}]  \left( \frac{1}{2}\Gamma^{(\alpha)}_2 \hat c^{(\alpha)}_2 + \sqrt{\Gamma^{(\alpha)}_2}\hat b^{(\alpha)}_{\text{in},2}   \right)
				+ \left( \frac{1}{2}\Gamma_2 \hat c^{(\alpha)\dagger}_2+ \sqrt{\Gamma^{(\alpha)}_2}\hat b_{\text{in},2}^{(\alpha)\dagger}   \right) [\hat a^{(\alpha)}_2,\hat c^{(\alpha)}_2].
			\end{align}
		\end{subequations}%
		Following the same procedure outlined in the previous section, we now recast the quantum Langevin equations for this cascaded setup into a form suitable for deriving a Lindblad-type master equation.
		We start by examining the output fields of each target, given by:
		\begin{equation}
			\hat b^{(\alpha)}_{\text{out},2}=\hat b^{(\alpha)}_{\text{in},2}+ \sqrt{\Gamma^{(\alpha)}_2}\hat c^{(\alpha)}_2= \frac{1}{\sqrt{2}}[\hat b_{\text{in}}+\hat b^{(\alpha)}_0]+ \sqrt{\frac{\Gamma_1}{2}}\hat c_1+ \sqrt{\Gamma_2}\hat c^{(\alpha)}_2 ,
		\end{equation}
		which suggests that the effective coupling in each arm can be described by a collective operator: $\hat C^{(\alpha)}\equiv \sqrt{\Gamma_1/2}\hat c_1+\sqrt{\Gamma^{(\alpha)}_2}\hat c^{(\alpha)}_2$.
		Using this, we can now write the full quantum Langevin equation governing the dynamics of an arbitrary system operator $\hat a$ as:
		\begin{multline}
			\dot {\hat a}=-i \left[\hat a,\hat H+\sum_{\alpha=1}^2 \frac{i}{2} 	\sqrt{\frac{\Gamma_1 \Gamma^{(\alpha)}_2}{2}} (\hat c_1^\dagger \hat  c^{(\alpha)}_2-\hat  c^{(\alpha)\dagger}_2 \hat  c_1) \right]
			+ \sum_{\alpha=1}^2 \sqrt{\frac{ \Gamma^{(\alpha)}_2}{2}} \left[(b^{(\alpha)\dagger}_0-\hat  b^\dagger_{\text{in}})[\hat  a,\hat c^{(\alpha)}_2]-[\hat a,\hat  c^{(\alpha)\dagger}_2](\hat  b^{(\alpha)}_0-\hat  b_{\text{in}})\right]\\
			+\sum_{\alpha=1}^2 \left[ -[\hat  a,\hat  C^{(\alpha)\dagger}]  \left( \frac{1}{2}\hat  C^{(\alpha)} + \sqrt{2 }\hat  b_{\text{in}}   \right)
			+ \left( \frac{1}{2}  \hat  C^{(\alpha)\dagger} + \sqrt{2 } \hat  b^\dagger_{\text{in}}   \right) [\hat  a,\hat  C^{(\alpha)}]  \right].
			\label{eq:LangevinEqTwo}
		\end{multline}
		Taking the expectation value of this equation and applying the quantum Ito calculus---as introduced in Eq.~\eqref{eq:ItoAlgebra}---, we arrive at the corresponding master equation for the full cascaded system:
		\begin{equation}
			\frac{d \rho(t)}{d t}=-i [\hat H,\hat \rho] 
			+\frac{\Gamma_1}{2}\mathcal{D}[\hat c_1]\hat \rho
			+ \sum_{\alpha=1}^2 \frac{\Gamma_2^{(\alpha)}}{2}\mathcal{D}[\hat c_2^{(\alpha)}]\hat \rho
			- \sum_{\alpha=1}^2\sqrt{\Gamma_1 \Gamma_2^{(\alpha)}/2}\left( [\hat c_2^{(\alpha)\dagger} , \hat c_1 \hat \rho]+ [\hat \rho \hat c_1^\dagger, \hat c_2^{(\alpha)}] \right).
			\label{eq:TwosensorCascadedMaster}
		\end{equation}
		We observe from Eq~\eqref{eq:TwosensorCascadedMaster} that is formally the same as Eq.~\eqref{eq:cascadedmaster2} by identifying:
		$\hat H \rightarrow \hat H + \Delta_1 \hat \xi^\dagger_1 \hat \xi_1+\Delta_2 \hat \xi^\dagger_2 \hat \xi_2$, $\hat c_1\rightarrow \hat \sigma$,  $\hat c_2^{(1)}\rightarrow \hat \xi_1$, $\hat c_2^{(2)}\rightarrow \hat \xi_2$, $\Gamma_1 \rightarrow \gamma$,   $\Gamma_{2}^{(1/2)}\rightarrow \Gamma$, and $\sqrt{\Gamma_1\Gamma_2^{(\alpha)}/2} \rightarrow \sqrt{\varepsilon \gamma\Gamma/2}$.
		This formalism captures the essential physics of cascaded coupling mediated by a beam splitter and rigorously accounts for the associated vacuum fluctuations. 
	\end{widetext}

	\section{Method to compute the steady-state and its differentiation}
	\label{Appendix:method_SS}
	
	In this section, we present the methods used to efficiently compute both the steady state and its derivative with respect to a system parameter, as the metrological quantities evaluated throughout this work depend explicitly on these calculations.

	\subsection{Computation of the steady-state}
	Let us assume an arbitrary open quantum system whose dynamics is described by a Lindblad master equation:
	\begin{equation}
		\frac{d \hat \rho}{dt }= -i[\hat H, \hat \rho]+ \sum_i \mathcal{D}[\hat L_i] \hat \rho,
	\end{equation}
	where $\hat H$ is the Hamiltonian, $\mathcal{D}[\hat L_i] $ is a general Lindblad term defined via the jump operator $\hat L_i$, and $\hat \rho$ is the reduced density matrix describing the quantum system.

	By casting the master equation in terms of a Liouvillian superoperator, $d\hat\rho/dt = \mathcal{\hat L} \hat \rho$, we can obtain the steady-state of the system by computing the nullspace of the Liouvillian,
	\begin{equation}
		\mathcal{\hat L} \hat \rho_{\text{ss}}=0,
		\label{eq:SteadyStateRelation}
	\end{equation}
	where the steady-state matrix is defined as $\hat \rho_{\text{ss}}\equiv\hat \Lambda_1^R$. This matrix corresponds to the right eigenmatrix of the Liouvillian, $\hat \Lambda_1^R$ associated with the zero eigenvalue, $\lambda_1 = 0$, as guaranteed by Evans’ theorem~\cite{EvansIrreducibleQuantum1977,EvansGeneratorsPositive1979}. However, the diagonalization of the Liouvillian may be computationally demanding, especially for large Hilbert spaces.

	We can implement a more efficient method to compute steady-states by considering the physical constraint of trace preservation: $\text{Tr}[\hat \rho_{\text{ss}}]=1$.  This constraint allows us to reduced the computational complexity by replacing, for instance, the first row of the Liouvillian with another with $1$s in the elements 
	\begin{equation}
		\mathcal{\hat L}_{1,i}= \delta_{1,i} \quad \text{when} \quad i= d (\alpha-1)+\alpha \quad \text{for} \quad \alpha=1,2, \ldots d,
	\end{equation}
	where $d$ is the dimension of the Hilbert space. That is, there will be a $1$ in those entries that multiply an element of the diagonal of the density matrix, and $0$ otherwise. This yields an effective Liouvillian matrix, $\mathcal{\hat L}_{\text{eff}}$, satisfying the following simple linear equation,
	\begin{equation}
		\mathcal{\hat L}_{\text{eff}} \hat \rho_{\text{ss}}=\mathbf{c} \quad \text{with}\quad \mathbf{c}=(1,0,\ldots,0)^T,
	\end{equation}
	such that the steady-state is simply given by
	\begin{equation}
		\hat \rho_{\text{ss}}=	\mathcal{\hat L}_{\text{eff}}^{-1} \mathbf{c}.
		\label{eq:FormalSteadyState}
	\end{equation}

	\subsection{Computation of the differentiation of the steady-state} 
	
	The formal relation from Eq.~\eqref{eq:FormalSteadyState} can be extended to compute the differentiation of the steady-state density matrix with respect to a system parameter $\theta$, denoted as $\partial_\theta \hat \rho_{\text{ss}}$. 
	This is achieved by taking the derivative with respect $\theta$ in Eq.~\eqref{eq:SteadyStateRelation}, such that
	\begin{equation}
		\partial_\theta [\mathcal{\hat L} \hat \rho_{\text{ss}}]= (\partial_\theta  \mathcal{\hat L} ) \hat \rho_{\text{ss}}+ \mathcal{\hat L} \partial_\theta \hat \rho_{\text{ss}}=0, 
	\end{equation}
	and get the relation
	\begin{equation}
		\partial_\theta \hat \rho_{\text{ss}}=-  \mathcal{\hat L}^{-1}(\partial_\theta  \mathcal{\hat L} ) \hat \rho_{\text{ss}}.
	\end{equation}
	Analogously to the steady-state computation, we can reduce the computational complexity by applying the preservation of the trace, and obtain a similar relation but in terms of the effective Liouvillian superoperator:
	\begin{equation}
		\partial_\theta \hat \rho_{\text{ss}}=-  \mathcal{\hat L}_{\text{eff}}^{-1}(\partial_\theta  \mathcal{\hat L}_{\text{eff}} ) \hat \rho_{\text{ss}}.
		\label{eq:DiffSteadyStateRelation}
	\end{equation}
	Note that before computing the differentiation of the steady state with respect to a parameter $\theta$, the steady state itself must first be determined. Using the formal relation for computing the steady-state in Eq.~\eqref{eq:SteadyStateRelation}, we can rewrite Eq.~\eqref{eq:DiffSteadyStateRelation} solely in terms of the effective Liouvillian,
	\begin{equation}
		\partial_\theta \hat \rho_{\text{ss}}=-  \mathcal{\hat L}_{\text{eff}}^{-1}(\partial_\theta  \mathcal{\hat L}_{\text{eff}} )	\mathcal{\hat L}_{\text{eff}}^{-1} \mathbf{c}\quad \text{with}\quad \mathbf{c}=(1,0,\ldots,0)^T.
		\label{eq:DiffSteadyStateRelation1}
	\end{equation}

	\begin{figure}[b!]
		\centering
		\includegraphics[width=0.88\linewidth]{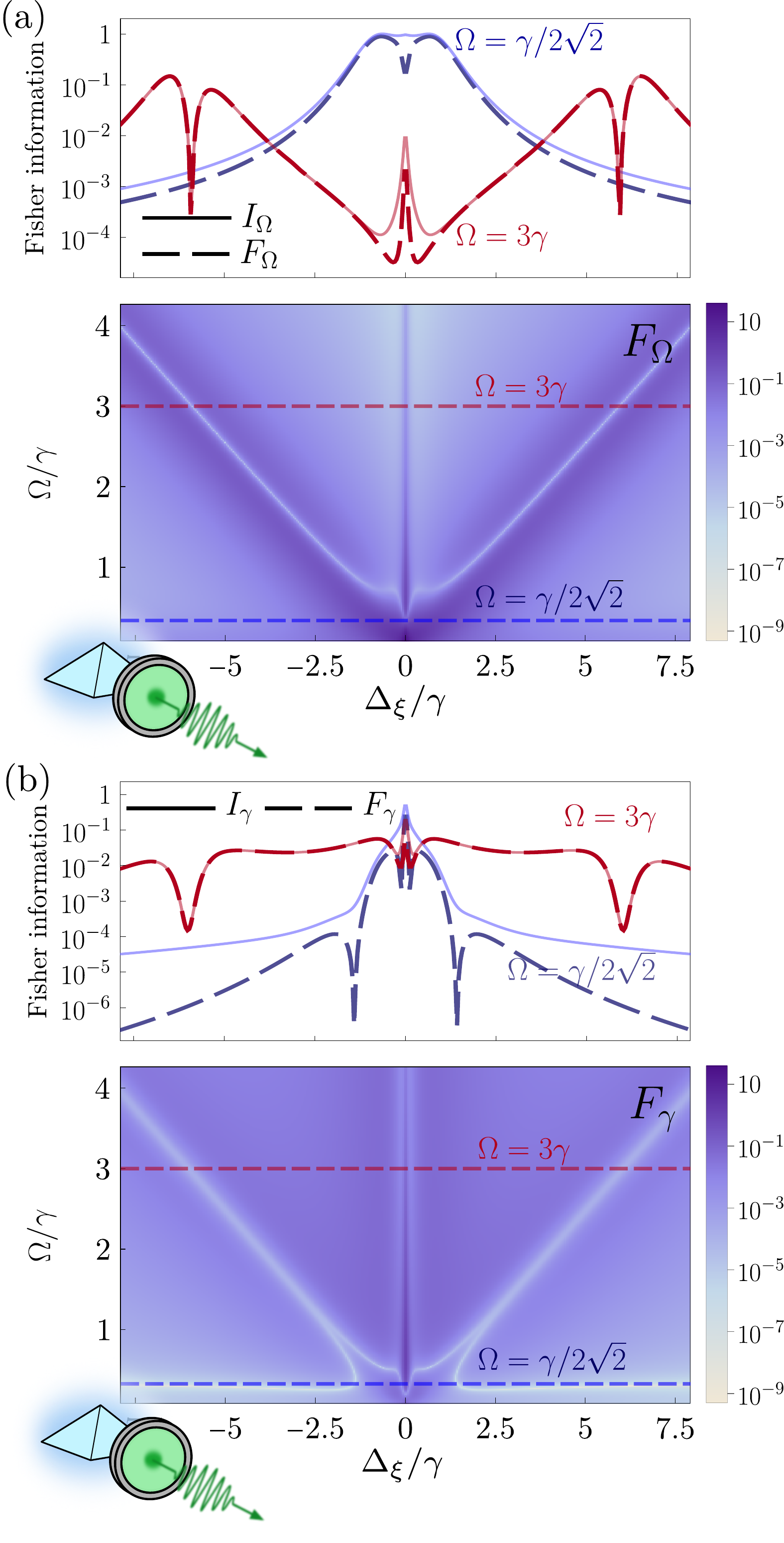}
		\caption{
			Frequency-resolved Fisher information for one-sensor to estimate (a) $\theta=\Omega$ and (b) $\theta=\gamma$.
			The top panels of each figure shows the classical Fisher information $F_\theta$ (dashed curves)  and the quantum Fisher information $I_\theta$ (solid curves) for two values of $\Omega=\{\gamma/2\sqrt{2}, 3\gamma \}$, in blue and red, respectively.
			The lower panels depict $F_\theta$ in terms of the driving strength $\Omega$ and the cavity frequency $\Delta_\xi$.
			Parameters: (a) $\Omega=3\gamma$; (a-c) $\Delta=10^{-6}\gamma$, $\Gamma=10^{-1}\gamma$, $\varepsilon=0.1$.
		}
		\label{fig:Appendix_OneSensor_FisherOmegaGamma}
	\end{figure}

	\section{One-sensor frequency-resolved Fisher information for $\theta\in \{\Omega,\gamma\}$}
	\label{Appendix:One_sensor_Other_parameters}

	In this section, we extend the analysis presented in Sec.~\ref{sec:sec3}, particularly the results in Fig.~\ref{fig:Single-sensor-Fisher-DrivingFrequency}, to the estimation of the parameters $\theta=\{\Omega, \gamma\}$. 
	
	Figure~\ref{fig:Appendix_OneSensor_FisherOmegaGamma} shows the CFI, $F_\theta$, as a function of the sensor-laser detuning $\Delta_\xi$ (top panels) and of the driving strength $\Omega$ (bottom panels), with panel (a) corresponding to $\theta=\Omega$ and panel (b) to $\theta=\gamma$.
	Similar to the case discussed in the main text, the CFI for these parameters exhibits rich spectral features that closely resemble the Mollow triplet structure [cf. Fig.~\ref{fig:Single-sensor-Fisher-DrivingFrequency}(b)]. In fact, for these two particular cases, we can observe how the CFI features local minima around the Mollow resonances at $\Delta_\xi\approx \pm 2\Omega$. 
	For $\theta=\Omega$ [panel (a)], we find that photon-counting measurements with $\alpha=0$ (dashed curves) nearly saturate the QFI (solid curves) across the frequency domain, as illustrated for $\Omega=\{\gamma/2\sqrt{2},3\gamma\}$ in blue and red, respectively. In this case, standard photon-counting essentially constitutes the optimal measurement strategy.

	A similar behavior is observed for $\theta=\gamma$ [panel (b)], with the exception of the Heitler regime, where standard photon-counting deviates noticeably from the QFI [see the solid and dashed blue curves in the top panel of (b)].

	This behaviour is similar to what occurs for $\theta=\gamma$, except in the Heitler regime, where the standard photon-counting features a non-negligible deviation from the QFI [see solid and dashed blue curves in the top panel of (b)].
	
	Additionally, we also mention that, in comparison with the case $\theta=\Delta$, the CFI for estimating $\theta\in\{\Omega,\gamma\}$ exhibits much higher values, although the spectral structure remains in any case.

	\section{Dependence on the imperfect coupling parameter $\varepsilon$}
	\label{Appendix:ImperfectCoupling}
	
	\begin{figure}[b!]
		\centering
		\includegraphics[width=1\linewidth]{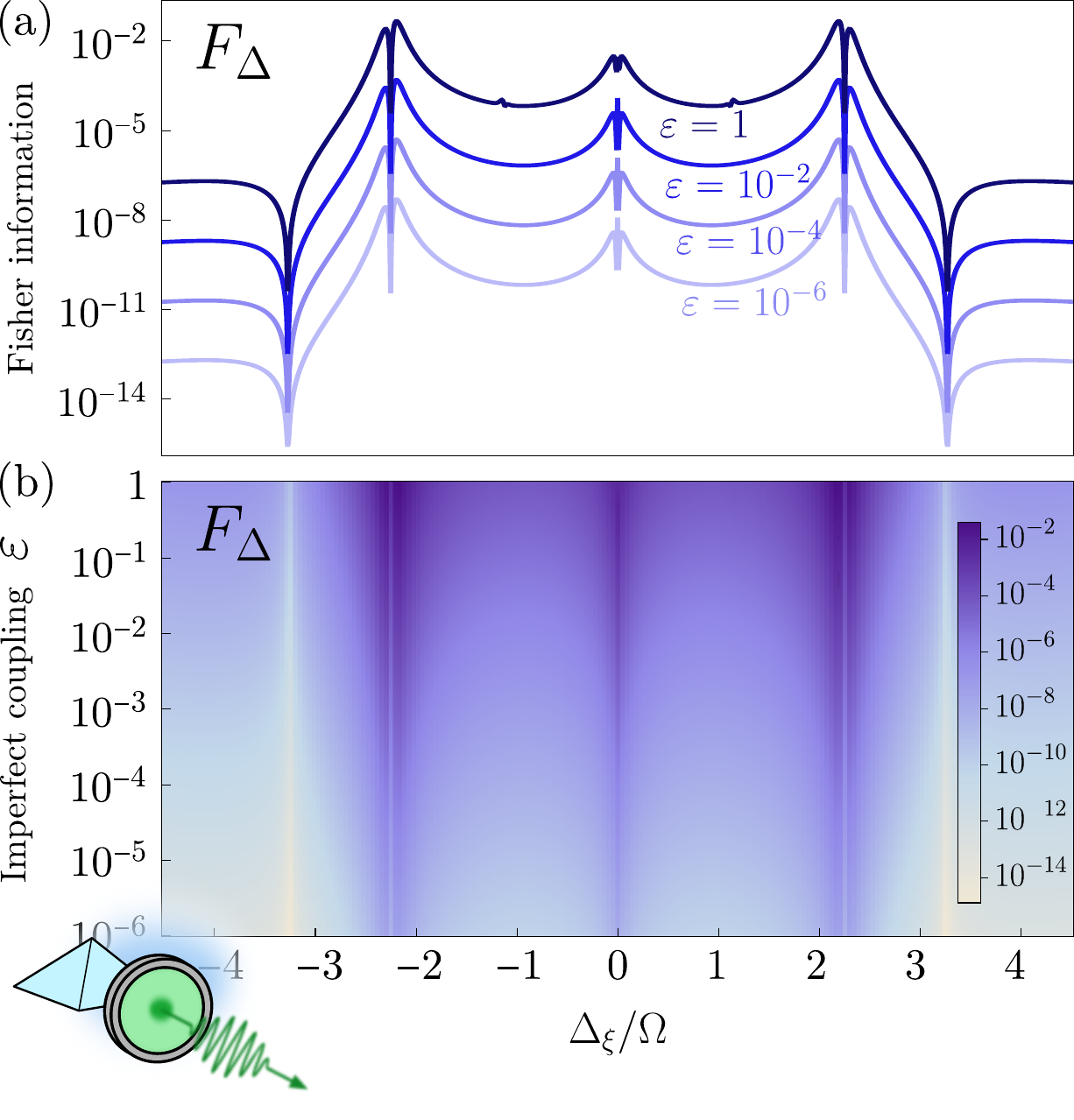}
		\caption{
			Impact of imperfect system-sensor coupling ($\varepsilon$) on the estimation sensitivity.
			(a)  $F_\Delta$ as a function of the sensor-laser detuning $\Delta_\xi$, shown for various coupling efficiencies $\varepsilon = \{10^{-6}, 10^{-4}, 10^{-2}, 1\}$, with lighter to darker curves representing increasing values of $\varepsilon$.
			(b) $F_\Delta$ as a function of both $\Delta_\xi$ and $\varepsilon$.
			Parameters: $\Delta = 0$, $\Omega = 10\gamma$, and $\Gamma = 0.1\gamma$.
		}
		\label{fig:ImperfectCoupling}
	\end{figure}

	In this section, we study how imperfect system-sensor coupling---characterized by the parameter $\varepsilon$---affects the classical Fisher information. For clarity and simplicity, we focus on the single-sensor scenario with no displacement ($\alpha = 0$) and use the expression of the CFI in terms of normally ordered correlators presented in Eq.~\eqref{eq:FisherInform_Corr_OneSensor}.

	According to the sensor method~\cite{DelValleTheoryFrequencyFiltered2012}, the stationary filtered correlators, $G^{(k)}_\Gamma=\langle\hat \xi^{k \dagger }\hat \xi^k \rangle$, are connected to the Fourier transform of the unfiltered $k$th autocorrelation function, 
	\begin{multline}
		S^{(k)}_\Gamma=(\Gamma/2\pi)^k\int_{\mathbb{R}}dt_1 dt_2 \ldots dt_k e^{\Gamma t_1/2} e^{\Gamma t_2/2}\ldots e^{\Gamma t_k/2}\\ \times\langle \hat a^{\dagger}(t_1) \hat a^{\dagger}(t_2)\ldots\hat a(t_2)\hat a(t_1)\rangle,
	\end{multline}
	by
	\begin{equation}
		G^{(k)}_\Gamma=\frac{g^{2k}}{\Gamma^k}(2\pi)^k S^{(k)}_\Gamma,
	\end{equation}
	where $g=\sqrt{\varepsilon \gamma\Gamma}$ is the cascaded system-sensor coupling.
	This relation reveals that a filtered correlator for two different imperfect couplings, $G^{(k)}_{\Gamma,\varepsilon_1}$ and $G^{(k)}_{\Gamma,\varepsilon_2}$, scale as
	\begin{equation}
		G^{(k)}_{\Gamma,\varepsilon_1}=\left( \frac{\varepsilon_1}{\varepsilon_2}\right)^{2k} G^{(k)}_{\Gamma,\varepsilon_2}.
		\label{eq:scaling_epsilon}
	\end{equation}
	Considering $\varepsilon_2=1$ (perfect coupling), we observe that the strength of $k$th order correlators is increasingly suppressed with decreasing $\varepsilon_1$, due to the exponential dependence on the order $2k$. Consequently, the estimation sensitivity---quantified by the CFI---is expected to deteriorate significantly under imperfect coupling.

	Substituting the relation Eq.~\eqref{eq:scaling_epsilon} into Eq.~\eqref{eq:FisherInform_Corr_OneSensor}, the CFI under imperfect coupling becomes:
	\begin{equation}
		F^\alpha_{\theta,\text{imperfect}}=
		\sum_{n}^{n_{\text{exc}}} \frac{
			\left[ \sum_{k\geq n}^{n_{\text{exc}}} c_{n,k} \varepsilon^{2k}  \partial_\theta G^{(k)}_{\Gamma,\text{perfect}} \right]^2
		}{
			\sum_{k\geq n}^{n_{\text{exc}}}  c_{n,k} \varepsilon^{2k} G^{(k)}_{\Gamma,\text{perfect}}
		}.
	\end{equation}
	This expression clearly illustrates the nonlinear degradation of CFI with decreasing $\varepsilon$. In particular, the contributions from higher-order correlations are exponentially suppressed, which can severely limit the sensitivity of the protocol, especially when relying on non-classical or multiphoton features of the light field.

	This dependence is illustrated in Fig.~\ref{fig:ImperfectCoupling}. In panel (a), we show the behavior of $F_\Delta$ as a function of detector frequency $\Delta_\xi$ for several values of the imperfect coupling parameter, $\varepsilon = \{10^{-6}, 10^{-4}, 10^{-2}, 1\}$. Panel (b) displays the joint dependence of $F_\Delta$ on both $\Delta_\xi$ and $\varepsilon$.
	As expected, the CFI decreases significantly for small values of $\varepsilon$, highlighting the crucial role of efficient system-sensor coupling in frequency-resolved quantum metrology. For instance, we observe that $F_\Delta$ increases from approximately $\sim 10^{-14}$ at $\varepsilon \approx 10^{-6}$ to $\sim 10^{-2}$ for perfect coupling ($\varepsilon = 1$).
	Interestingly, despite the substantial change in magnitude, the spectral structure of the CFI remains qualitatively unchanged across all values of $\varepsilon$. This robustness underscores a key strength of the method: even under extremely weak coupling, the underlying spectral features of the emitted light remain accessible to detection. Furthermore, we emphasize that the sensitivity can still be improved through mean-field engineering techniques, which enable a systematic and robust strategy to enhance the Fisher information, even under imperfect coupling conditions.

	\section{Poissonian Fisher information}
	\label{Appendix:Poissonian_Fisher}
	
	In this section, we derive the Poissonian Fisher information $F_\theta^P$ presented in Eq.~\eqref{eq:PoissonianFisher}, and establish a bound comparing it with the classical Fisher information.

	\subsection{Derivation}
	
	Following the same approach that is used in Refs.~\cite{DelaubertQuantumLimits2008,Vivas-VianaTwophotonResonance2021}, we assume that the resulting state of the sensor is a
	coherent state, so that the corresponding measured photo-current displays
	Poissonian fluctuations. Under this assumption, for a sensor at frequency $\omega_\xi$, the photon-counting distribution is given by:
	\begin{equation}
		p(n|\theta)=\frac{\bar n^n e^{-\bar n}}{n!},
	\end{equation}
	where $n$ denotes the excitation number and $\bar n=\langle \hat \xi^\dagger \hat \xi \rangle$, i.e., the mean photon distribution conditioned on the parameter $\theta$.
	By inserting this conditional probability distribution into the general definition of the classical Fisher information in Eq.~\eqref{eq:CFI_PC}, and using the identity
	$
	\partial_\theta p(n|\theta)=p(n|\theta) \left(\frac{n}{\bar n}-1 \right)(\partial_\theta \bar n)$, along with the moments $\sum_n n p(n|\theta)=\bar n$ and $\sum_n n^2 p(n|\theta)= \bar n^2+\bar n$, we arrive at the expression for the Poissonian Fisher information [Eq.~\eqref{eq:PoissonianFisher}]:
	\begin{equation}
		F_\theta^P=\frac{[\partial_\theta \bar n]^2}{\bar n}=\frac{[\partial_\theta \langle \hat \xi^\dagger \hat \xi \rangle]^2}{\langle \hat \xi^\dagger \hat \xi \rangle}.
	\end{equation}

	\subsection{Poissonian Fisher information bound}
	
	Although it is intuitive to assume that a full statistical description of light yields higher sensitivity in metrological protocols, this is not mathematically guaranteed. In particular, the classical Fisher information (CFI) may not always outperform the Fisher information obtained under a Poissonian approximation.
	
	To illustrate this, we apply Sedrakyan’s inequality~\cite{SedrakyanAlgebraicInequalities2018}, which states that for real numbers $a_1,a_2, \ldots, a_n$ and strictly positive real numbers $b_1,b_2, \ldots, b_n$, the following inequality holds
	$$
	\sum_{k=1}^n \frac{a_k^2}{b_k} \geq \frac{(\sum_{k=1}^n a_k)^2}{\sum_{k=1}^n b_k}.
	$$
	This inequality is essentially a reformulation of the Cauchy–Schwarz inequality (CSI), which reads:
	\begin{equation}
		\left(\sum_{k=1}^n x_k^2 \right) \left(\sum_{k=1}^n y_k^2 \right)\geq \left(\sum_{k=1}^n x_k y_k \right)^2,
	\end{equation}
	which can be derived by setting $x_k=a_k/\sqrt{b_k}$ and $y_k=\sqrt{b_k}$. Under these definitions, the CSI simplifies to the Sedrakyan's inequality.
	We now apply this inequality to the definition of the Poissonian Fisher information, Eq.~\eqref{eq:PoissonianFisher}:
	\begin{equation}
		F_\theta^P =\frac{[\sum_n n  \partial_\theta  p(n|\theta)]^2}{\sum_n n p(n|\theta)} \leq \sum_n n \frac{[\partial_\theta p(n|\theta)]^2}{p(n|\theta)}=\sum_n n F^{(n)},	
		\label{eq:cond1}
	\end{equation}
	where we can define $F^{(n)}\equiv [\partial_\theta p(n|\theta)]^2/p(n|\theta)$ as the $n$th order term in the expansion of the Fisher information, $F_\theta= \sum_n [\partial_\theta p(n|\theta)]^2/p(n|\theta)= \sum_n F^{(n)}$. 
	Noting that $n F^{(n)} \geq F^{(n)}$ since $n\in \mathbb{N}^+$, we also obtain the inequality
	\begin{equation}
		F_ \theta= F^{(0)}+ \sum_{n=1} F^{(n)} \leq F^{(0)} + \sum_{n=1} n F^{(n)}. 
		\label{eq:cond2}
	\end{equation} 
	Considering these two conditions, \cref{eq:cond1,eq:cond2}, we arrive at the bound 
	\begin{equation}
		|F_\theta - F_\theta^P| \leq F^{(0)}.
	\end{equation}
	Hence, the intuitive notion that $F_\theta$ must always be greater than $F_\theta^P$ fails, since it may occur that the variance of a given operator, along with a general quantum probability distribution, is greater than the one associated to a Poissonian probability distribution, and thus it may occur $F_ \theta<F_\theta^P$.
	Therefore, in practical metrological scenarios, it may be more reliable to benchmark performance using the signal-to-noise ratio (SNR), which obeys the mathematical bound:
	$F_\theta \geq \mathrm{SNR}_\theta$, and thus serves as a consistent and theoretically grounded figure of merit.

	\section{Mean-field engineering}
	\label{Appendix:MeanField}
	
	In this section, we provide further details on the mean-field engineering technique used in the main text. Specifically, we describe its mathematical implementation for both single- and two-sensor cases via coherent displacements in phase space using Laguerre polynomials. 
	We also show how to isolate the quantum fluctuations by canceling out the coherent contribution of the signal.
	
	\subsection{General method}
	
	As explained in the main text [see Fig.~\ref{fig:MF-engineering}(a)], the mean-field engineering technique is implemented by means of an unbalanced beam splitter combined with a local oscillator in the limit of perfect transmittance. 
	In this scenario, the sensor receives the output field displaced by a complex amplitude $\alpha$ in phase space. 
	
	Formally, this process corresponds to displacing the sensor density matrix via
	\begin{equation}
		\hat \rho_\theta^{\alpha}=\hat D(\alpha) \hat \rho_\theta \hat  D^\dagger(\alpha),
	\end{equation}
	where $\hat \rho_\theta$ is the density matrix of the sensor, and $\hat  D(\alpha)=\exp(\alpha \hat \xi^\dagger - \alpha^*\hat \xi)$ is the displacement operator. 
	Projecting onto the Fock basis, $\langle n|\cdot|n\rangle$, yields the generalized photon-counting POVM introduced in Eq.~\eqref{eq:POVMs}.

	Since our analysis focuses on photon-counting statistics, only the diagonal elements of the displaced density matrix are relevant.
	Consequently, it is unnecessary to compute the full transformation of the quantum state. Instead, the transformation simplifies to the evaluation of the diagonal elements:
	\begin{equation}
		p^{\alpha}(n|\theta)=\langle n| D(\alpha) \hat \rho_\theta D(\alpha)^\dagger |n\rangle.
	\end{equation}
	Now, using the general decomposition of a density matrix, $\hat \rho_\theta=\sum_{\mu,\nu}\rho_\theta^{\mu,\nu} |\mu \rangle \langle \nu |$, with $\rho_\theta^{\mu,\nu}\equiv \langle \mu |\hat  \rho_\theta |\nu\rangle$, the displaced probability distribution becomes
	\begin{equation}
		p^{\alpha}(n|\theta)=\sum_{\mu,\nu}  \rho_\theta^{\mu,\nu} \langle n| D(\alpha) |\mu \rangle \langle \nu |  D(\alpha)^\dagger |n\rangle.
	\end{equation}
	The matrix elements of the displacement operator in the Fock basis are easily given by the identity~\cite{CahillDensityOperators1969,DeOliveiraPropertiesDisplaced1990}:
	\begin{equation}
		\langle m | \hat D(\alpha)|n\rangle= \sqrt{\frac{n!}{m!}} \alpha^{m-n} e^{-|\alpha|^2/2}  L_n^{(m-n)} (|\alpha|^2) 
		\quad (m\geq n),
	\end{equation}
	where $L_n^{(m-n)} (x)$ is an associated Laguerre polynomial~\cite{AbramowitzHandbookMathematical1972}. 
	As a result, using this identity, the displaced photon-counting distribution reduces to
	\begin{widetext}
		\begin{equation}
			p^\alpha(n|\theta) = \sum_{\mu,\nu} \rho_\theta^{\mu,\nu} \sqrt{\frac{\mu!}{\nu!}} (-1)^{\nu -n} \alpha^{\nu-\mu} e^{-|\alpha|^2}L_n^{(n-\mu)} (|\alpha|^2)   L_n^{(\nu-n)} (|\alpha|^2) \quad  (n\geq \mu, \nu \geq n).
		\end{equation}

		In the case of two sensors, we define the joint displacement operator as the product of two independent single-mode displacements:
		\begin{equation}
			\hat D(\alpha_1,\alpha_2)=\hat D_1(\alpha_1)\hat D_2(\alpha_2)=e^{\alpha_1 \hat \xi_1^\dagger +\alpha_2 \hat \xi_2^\dagger-\alpha_1^* \hat \xi_1 -\alpha_2^* \hat \xi_2 },
		\end{equation}
		with each sensor possibly displaced by a different amplitude $\alpha_1$ and $\alpha_2$.
		Then, the displaced two-sensor density matrix is given by 
		\begin{equation}
			\hat \rho_\theta^{\alpha_1,\alpha_2}=\hat D(\alpha_1,\alpha_2) \hat \rho_\theta \hat  D^\dagger(\alpha_1,\alpha_2),
		\end{equation}
		Taking the diagonal projection in the two-mode Fock basis, $\langle n_1 n_2| \cdot |n_1n_2\rangle$, yields the joint photon-counting probability:
		\begin{equation}
			p^{\alpha_1,\alpha_2}(n_1,n_2|\theta) = \langle n_1 n_2 | \hat D(\alpha_1,\alpha_2) \hat \rho_\theta \hat D^\dagger(\alpha_1,\alpha_2) | n_1 n_2 \rangle.
		\end{equation}
		Using the displacement matrix elements for two modes:

		\begin{multline}
			\langle m_1, m_2 | \hat D(\alpha_1,\alpha_2)|n_1, n_2\rangle= \sqrt{\frac{n_1!n_2!}{m_1!m_2!}} \alpha_1^{m_1-n_1}\alpha_2^{m_2-n_2} e^{-|\alpha_1|^2/2}e^{-|\alpha_2|^2/2} \\
			\times L_{n_1}^{(m_1-n_1)} (|\alpha_1|^2) L_{n_2}^{(m_2-n_2)} (|\alpha_2|^2) 
			\quad (m_1,m_2\geq n_1,n_2),
			\label{eq:TwoSensorsDispOveralp}
		\end{multline}
		the joint probability distribution reduces to: 
		\begin{multline}
			p^{\alpha_1,\alpha_2}(n_1,n_2|\theta) = \sum_{\mu,\nu}\sum_{\mu',\nu'} \rho_\theta^{\mu\nu,\mu'\nu'} \sqrt{\frac{\mu!\nu!}{\mu'!\nu'!}} (-1)^{(\mu'+\nu') -(n_1+n_2)} \alpha_1^{n_1-\mu}{\alpha_1^*}^{\mu'-n_1}\alpha_2^{n_2-\nu}{\alpha_2^*}^{\nu'-n_2} e^{-|\alpha_1|^2-|\alpha_2|^2} \\
			\times L_\mu^{(n_1-\mu)} (|\alpha_1|^2)   L_\nu^{(n_2-\nu)}(|\alpha_2|^2)
			L_{n_1}^{(\mu'-n_1)} (|\alpha_1|^2)   L_{n_2}^{(\nu'-n_2)}(|\alpha_2|^2) \quad  (n_1,n_2\geq \mu,\nu; \mu'\nu' \geq n_1,n_2),
		\end{multline}
		where we have used the decomposition $\hat \rho_\theta = \sum_{\mu,\nu} \sum_{\mu’,\nu’} \rho_\theta^{\mu\nu,\mu’\nu’} |\mu\nu\rangle \langle \mu’\nu’|$, with coefficients $\rho_\theta^{\mu\nu,\mu’\nu’} \equiv \langle \mu\nu | \hat \rho_\theta | \mu’\nu’ \rangle$.

	\end{widetext}
	
	\subsection{Isolating quantum fluctuations}
	
	In the previous section, we described how to coherently displace a sensor state via a classical field to implement mean-field engineering. Here, we determine the specific value of the displacement parameter $\alpha$ that cancels the coherent component of the signal, isolating only the quantum fluctuations. For practical purposes, here we prove this using the single-sensor scenario as the extension to the two-sensor case is straightforward.
	
	We begin by noting that the TLS operator $\hat \sigma$ can be linearized into a mean-field and a fluctuation component,
	\begin{equation}
		\hat \sigma =\delta\hat \sigma+\langle \hat \sigma \rangle,
	\end{equation}
	where $\langle \delta \hat \sigma \rangle = 0$ by definition.
	Substituting this decomposition into the single-sensor cascaded master equation [Eq.~\eqref{eq:cascadedmaster1}], we obtain a modified Hamiltonian:
	\begin{multline}
		\hat H\rightarrow\Delta (\delta \hat \sigma^\dagger \delta \hat \sigma+|\langle \hat \sigma \rangle|^2 +\langle \hat \sigma\rangle\delta \hat \sigma^\dagger+\langle \hat \sigma\rangle^*\delta \hat \sigma) \\
		+\Omega(\delta\hat \sigma+\delta\hat \sigma^\dagger +\langle \hat \sigma\rangle+\langle \hat \sigma\rangle^*)
	\end{multline}
	The corresponding Lindblad term becomes:
	\begin{equation}
		\frac{\gamma}{2}\mathcal{D}[\hat \sigma]\hat \rho \rightarrow   \frac{\gamma}{2}\mathcal{D}[\delta \hat \sigma]\hat \rho+\frac{\gamma}{2}[(\langle \hat \sigma\rangle^*\delta\hat \sigma-\langle \hat \sigma\rangle\delta\hat \sigma^\dagger),\hat \rho],
	\end{equation}
	and the cascaded coupling transforms into:
	\begin{multline}
		\left\{ [\hat \xi^\dagger, \hat \sigma  \hat \rho ] + [\hat \rho \hat \sigma^\dagger, \hat \xi] \right\} \rightarrow \left\{ [\hat \xi^\dagger, \delta\hat \sigma  \hat \rho ] + [\hat \rho \delta \hat  \sigma^\dagger, \hat \xi] \right\} \\
		+ [(\langle \hat \sigma\rangle^*\hat \xi-\langle \hat \sigma\rangle\hat \xi^\dagger),\hat \rho].
	\end{multline}
	Collecting these terms, the transformed master equation becomes:
	\begin{multline}
		\frac{d \hat \rho }{dt}= -i [\tilde H+ \Delta_\xi \hat \xi^\dagger \hat \xi, \hat \rho] + \frac{\gamma}{2} \mathcal{D}[\delta\hat \sigma] \hat \rho 
		+  \frac{\Gamma}{2} \mathcal{D}[\hat \xi]\hat \rho \\
		-
		\sqrt{\varepsilon \gamma \Gamma} \left\{ [\hat \xi^\dagger, \delta\hat \sigma  \hat \rho ] + [\hat \rho \delta \hat \sigma^\dagger, \hat \xi] \right\},
	\end{multline}
	where we have defined an effective Hamiltonian, $\tilde H$,
	\begin{multline}
		\tilde H= \Delta (\delta \hat \sigma^\dagger \delta \hat \sigma+\langle \hat \sigma\rangle\delta \hat \sigma^\dagger+\langle \hat \sigma\rangle^*\delta \hat \sigma) 
		+\Omega(\delta\hat \sigma+\delta\hat \sigma^\dagger)
		\\
		+i\frac{\gamma}{2}(\langle \hat \sigma\rangle^*\delta\hat \sigma-\langle \hat \sigma\rangle\delta\hat \sigma^\dagger)
		+i \sqrt{\varepsilon \gamma \Gamma} (\langle \hat \sigma\rangle^*\hat \xi-\langle \hat \sigma\rangle\hat \xi^\dagger).
	\end{multline}
	The last two terms in the effective Hamiltonian represent coherent driving fields acting on the TLS and the bosonic sensor, respectively.
	However, due to the cascaded structure of the system, only the final term directly influences the sensor. This term can be interpreted as an effective coherent displacement induced by the coherent component of the emission from the source.
	Therefore, in order to remove this coherent contribution, we must apply a coherent displacement, 
	\begin{equation}
		\hat D(\alpha)=e^{\alpha\hat \xi-\alpha^*\hat \xi^\dagger}.
	\end{equation}
	Now, by considering the displacement of bosonic modes as $\hat D^\dagger (\alpha)\hat \xi^{(\dagger)} \hat D(\alpha)=\hat \xi+\alpha^{(*)}$, such that the Hamiltonian of the sensor becomes
	\begin{multline}
		\hat H_{\text{sensor}}=\Delta_\xi (\hat \xi^\dagger \hat \xi+ |\alpha|^2+\alpha^* \hat \xi +\alpha \hat \xi^\dagger) 
		\\
		+i\sqrt{\varepsilon\gamma \Gamma} [\langle \hat \sigma\rangle^* (\hat \xi +\alpha)+\langle \hat \sigma\rangle (\hat \xi^\dagger +\alpha^*)]+
		i\frac{\Gamma}{2}(\alpha^* \hat \xi-\alpha \hat \xi^\dagger),
	\end{multline}
	where the last term originates the displacement of the Lindblad term.
	Then, in order to remove the coherent contribution in the sensor Hamiltonian, $\hat H_{\text{sensor}}$, we must find the solution to the condition
	\begin{equation}
		\hat \xi \left( \Delta_\xi \alpha^*+i\sqrt{\varepsilon \gamma \Gamma}\langle{\hat \sigma\rangle^*+i\frac{\Gamma}{2}} \alpha^* \right)=0.
	\end{equation}
	From this, we get the value of $\alpha_{\text{fluct}}$ that cancels the coherent contribution, given by
	\begin{equation}
		\alpha_{\text{fluct}}=\frac{\sqrt{\varepsilon\gamma \Gamma}\langle \hat \sigma\rangle}{\Gamma/2+i\Delta_\xi}.
	\end{equation}
	Since we are considering stationary measurements, we can easily obtain the expression for $\langle \hat \sigma\rangle_{\text{ss}}$ from the source master equation in Eq.~\eqref{eq:MasterEqTLS},
	\begin{equation}
		\langle \hat \sigma\rangle_{\text{ss}}=-\frac{2i\gamma\Omega+4\Delta\Omega}{\gamma^2+4\Delta^2+8\Omega^2},
	\end{equation}
	such that the sensor density matrix of the quantum fluctuations is then given by
	\begin{equation}
		\hat \rho_{\text{fluct}}=\hat D^\dagger(\alpha_{\text{fluct}}) \hat \rho D(\alpha_{\text{fluct}}).
	\end{equation}
	
	\begin{figure}[t!]
		\centering
		\includegraphics[width=0.835\linewidth]{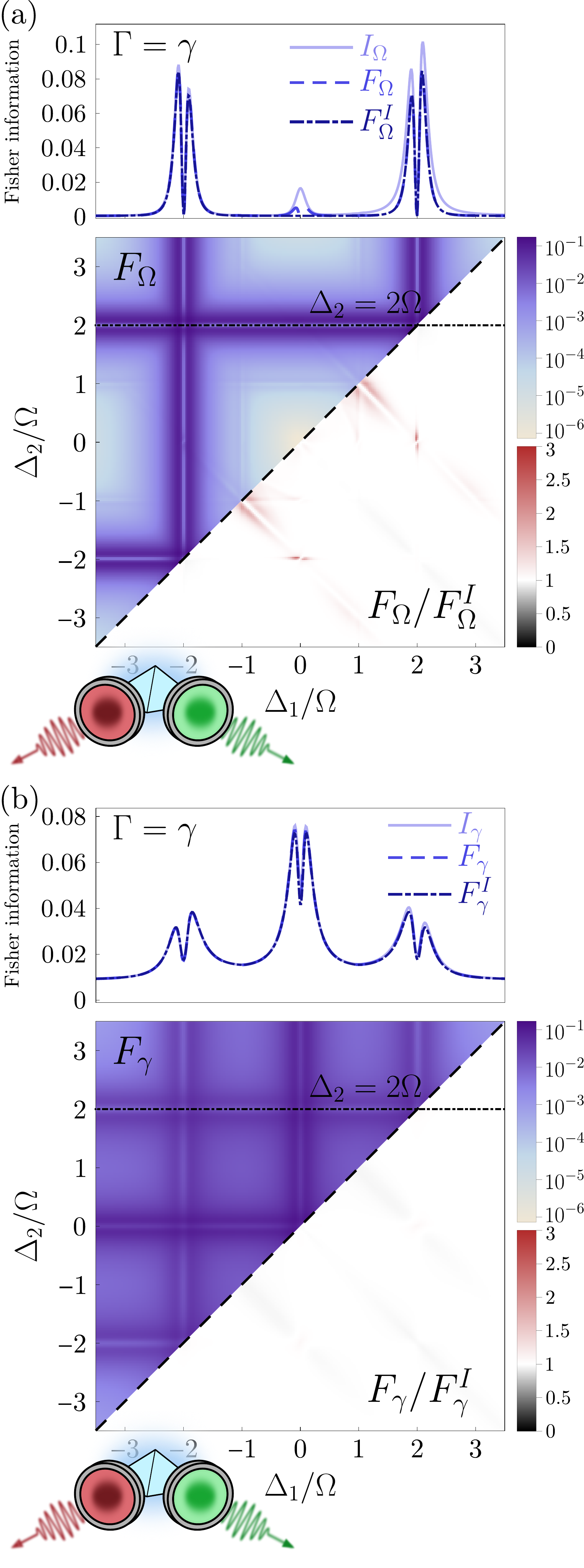}
		\caption{
			Frequency-resolved Fisher information for two-sensor to estimate (a) $\theta=\Omega$ and (b) $\theta=\gamma$.
			Top panels: $F_\theta$ (blue dashed curve), $F_\theta^I$ (blue dot-dashed curve), $I_\theta$ (blue solid curve) in terms of one sensor-laser detuning $\Delta_1$, with the second fixed at  $\Delta_2=2\Omega$. 
			Bottom panels: $F_\Delta$ (top left triangle) and $F_\theta/F_\theta^I$ (bottom right triangle) in the frequency domain ($\Delta_1,\Delta_2$).
			Parameters: (a-c) $\Omega=10\gamma,\ \Delta=0,\ \Gamma=\gamma,\ \varepsilon=0.5.$
		}
		\label{fig:TwoSensors_FisherOmegaGamma}
	\end{figure}
	
	\section{Two-sensor frequency resolved Fisher information for $\theta\in\{\Omega,\gamma\}$}
	\label{Appendix:Two_sensor_Other_parameters}

	In this section, we extend the results presented in Sec.~\ref{sec:sec4}, particularly in Fig.~\ref{fig:Two-sensor-CFI}, to the cases of $\theta\in \{\Omega, \gamma\}$. 

	The classical Fisher information for these two parameters, $\theta=\Omega$ [Fig.~\ref{fig:TwoSensors_FisherOmegaGamma}(a)] and $\theta=\gamma$ [Fig.~\ref{fig:TwoSensors_FisherOmegaGamma}(b)],  reaches significantly larger values than in the case $\theta=\Delta$.
	Notably, standard photon-counting measurements---i.e., $\vec \alpha=\vec 0$--already saturate the quantum Fisher information (QFI), indicating that no further measurement optimization is required to attain the ultimate precision limit, as we already observed for the case of a single sensor in Fig.~\ref{fig:Appendix_OneSensor_FisherOmegaGamma} [see Appendix~\ref{Appendix:One_sensor_Other_parameters}.].
	Moreover, when comparing the joint two-sensor CFI with its independent counterpart, $F_\theta^I$ [see Eq.~\eqref{eq:uncorrelatedBound}], we can observe that there is practically no metrological gain from retaining cross-correlations. As shown in the lower-right panels of Fig.~\ref{fig:TwoSensors_FisherOmegaGamma}, retaining cross-correlations provides at most a threefold improvement over independent measurements, in stark contrast with the case  $\theta=\Delta$ [see Fig.~\ref{fig:Two-sensor-CFI}], where cross-correlations yielded enhancements of up to five orders of magnitude near leapfrog resonances. 
	
	The study of these two parameters, $\theta\in \{\Omega,\gamma\}$, highlights that the metrological advantage of nonclassical correlations in the emitted photons strongly depends on the parameter of interest. While frequency-resolved correlations are crucial for estimating the qubit-laser detuning $\theta=\Delta$, these play a much less significant role for $\theta\in \{\Omega,\gamma\}$. That is, the sensitivity of frequency-resolved Fisher information is governed not only by the intrinsic photon statistics of the emission but also by how the relevant parameter is encoded in the output field.

	\section{Derivation of the joint probability distribution in terms of multimode correlators}
	\label{Appendix:ProbDistrbCorr}
	
	In this section, we derive the expression for the joint photon-number probability distribution in terms of normally ordered multimode correlators, as presented in Eq.~\eqref{eq:joint_prob_corr}. 
	This generalizes the approach used for a single mode---originally reported in Ref.~\cite{GardinerQuantumNoise2004} and later extended in Ref.~\cite{ZubizarretaCasalenguaStructureHarmonic2017}---to the case of joint detection involving two harmonic oscillators (sensors).

	\subsection{Proof}

	We consider two bosonic modes with Hilbert spaces, $\mathcal{H}_1$ and $\mathcal{H}_2$, both truncated up to $N+1$ excitations, such that the Hilbert space is $\mathcal{H} = \mathcal{H}_1 \otimes \mathcal{H}_2$.
	A general density matrix $\hat{\rho}$ in this space can be expanded in the Fock basis as:
	\begin{equation}
		\hat \rho = \sum_{n_1,n_2=0}^{N} P_{n_1,n_2} |n_1 n_2\rangle \langle n_1 n_2| + \text{off-diagonal terms},
	\end{equation}
	where $P_{n_1,n_2} \equiv \langle n_1 n_2 | \hat \rho | n_1 n_2 \rangle$ is the joint probability distribution of interest.
	To express this probability distribution in terms of correlators, as explained in Ref.~\cite{ZubizarretaCasalenguaStructureHarmonic2017}, we introduce the general moment:
	\begin{equation}
		G^{(\alpha,\beta)}\equiv \langle \hat \xi_1^{\dagger \alpha} \hat \xi_1^\alpha \hat \xi_2^{\dagger \beta}\hat \xi_2^\beta \rangle
	\end{equation}
	which, using the number-state representation, can be written as
	\begin{equation}
		G^{(\alpha,\beta)} = \sum_{n_1,n_2=0}^N \frac{n_1!}{(n_1-\alpha)!}\frac{n_2!}{(n_2-\beta)!}P_{n_1n_2}.
		\label{eq:Appendix_general_corr}
	\end{equation}
	This defines a linear system relating the vector of correlators $\vec{G}$ to the vector of probabilities $\vec{P}$, with a transformation matrix $\mathbb{M}$ encoding the combinatorial coefficients:
	\begin{equation}
		\vec G=\mathbb{M} \vec P,
	\end{equation}
	where
	\begin{subequations}
		\begin{align}
			\vec G&=(G^{(0,0)}, G^{(1,0)},\ldots,G^{(N,0)}, G^{(0,1)}, G^{(N,1)},\ldots, G^{(N,N)}), \\
			\vec P&=(P_{0,0}, P_{1,0},\ldots,P_{N,0}, P_{0,1}, P_{N,1},\ldots, P^{N,N}),
		\end{align}
	\end{subequations}
	while $\mathbb{M}$ is determined by Eq.~\eqref{eq:Appendix_general_corr}. 
	However, since the coefficient matrix $\mathbb{M}$ is not upper-triangular, a  direct inversion of this system is not possible, meaning that we cannot establish a clear bijection between probability distribution and correlators due to redundancies: the number of variables exceeds the number of equations. To solve this, we impose three physical constraints:
	\begin{enumerate}
		\item Normalization. The total probability must sum one,
		\begin{equation}
			\sum_{n_1,n_2=0}^N P_{n_1,n_2}=1
		\end{equation}
		\item Marginal distributions. The reduced single-mode distributions must agree with the marginals of $P_{n_1,n_2}$,
		\begin{subequations}
			\begin{align}
				P_{n_1}=\sum_ {n_2=0}^N P_{n_1,n_2} , \\ 
				P_{n_2}=\sum_ {n_1=0}^N P_{n_1,n_2},
			\end{align}
		\end{subequations}
		such that 
		\begin{subequations}
			\begin{align}
				P_{0,0}&=1- \sum_{n_1,n_2\neq 0}P_{n_1,n_2}\\
				P_{n_1,0}&=P_{n_1}-\sum_ {n_1\neq 0, n_2\neq 0}^N P_{n_1,n_2} , \\ 
				P_{0,n_2}&=P_{n_2}-\sum_ {n_1\neq 0, n_2\neq 0}^N P_{n_1,n_2}.
			\end{align}
		\end{subequations}
	\end{enumerate}
	Using these constraints, we eliminate redundant variables such as $P_{0,0}$, $P_{n_1,0}$, and $P_{0,n_2}$, reducing the number of variables from $(N+1)^2$ to $N^2$. 
	
	\begin{widetext}

		On this reduced space, the coefficient matrix becomes lower-triangular, allowing for an explicit recursive solution.
		These $N^2$ variables do not contain the zeros ($n_1,n_2=0$), so that now we can establish the bijection between correlator and probability distribution elements as the the linear system is triangular:
		\begin{equation}
			\vec G'=\mathbb{M}' \vec P',
		\end{equation}
		where
		\begin{subequations}
			\begin{align}
				\vec G'&=(G^{(1,1)},G^{(2,1)},\ldots,G^{(N,1)}, G^{(1,2)}, G^{(N,2)},\ldots, G^{(N,N)}), \\
				\vec P'&=(P_{1,1},P_{2,1},\ldots,P_{N,1}, P_{1,2}, P_{N,2},\ldots, P_{N,N}),
			\end{align}
		\end{subequations}
		and $\mathbb{M}'$ is the reduced coefficient matrix generated by $\vec G'$. As a consequence, we can now invert the relation of this linear system by solving it recursively (backward Gaussian substitution), such that
		\begin{equation}
			P_{n_1,n_2}=\sum_{j_1\geq n_1}^{n_{\text{exc}}} \sum_{j_2\geq n_2}^{n_{\text{exc}}}  \frac{(-1)^{n_1+j_1}}{n_1!(j_1-n_1)!} \frac{(-1)^{n_2+j_2}}{n_2!(j_2-n_2)!}
			G^{(j_1,j_2)},
		\end{equation}
		which expresses the joint photon-number distribution in terms of the normally ordered correlators, as we presented in Eq.~\eqref{eq:joint_prob_corr}.
		
		\subsection{Extension to $N$ bosonic modes}
		
		This procedure can be generalized to any number of modes. For an $N$-mode system ($\mathcal{H}=\bigotimes_{i=1}^N \mathcal{H}_i$), the joint distribution $P_{n_1,n_2,\ldots,n_N}$ can be reconstructed via a multi-dimensional extension of the same formula:
		%
		%
		%
		%
		%
		\begin{equation}
			P_{n_1,n_2,\ldots, n_N}=\sum_{j_1\geq n_1}^{n_{\text{exc}}} \sum_{j_2\geq n_2}^{n_{\text{exc}}} \ldots \sum_{j_N\geq n_N}^{n_{\text{exc}}}  \frac{(-1)^{n_1+j_1}}{n_1!(j_1-n_1)!} \frac{(-1)^{n_2+j_2}}{n_2!(j_2-n_2)!} \ldots \frac{(-1)^{n_N+j_N}}{n_N!(j_N-n_N)!}
			G^{(j_1,j_2,\ldots , j_N)}.
		\end{equation}
		This result provides the theoretical foundation for computing joint detection statistics from multimode quantum light using only a finite set of normally ordered moments.
		

		\section{Bound for the uncorrelated Fisher information}
		\label{Appendix:UncorrelatedBound}
		
		Here, in this section, we are going to prove the inequality presented in Eq.~\eqref{eq:uncorrelatedBound}, establishing that $F_\theta[p(n_1,n_2|\theta)]\geq 1/2(F_\theta[p(n_1|\theta)]+F_\theta[p(n_2|\theta)])$.
		To prove this, we consider the case of a general joint density matrix $\hat \rho_\theta$ describing the quantum state of both bosonic sensors. Although in practice the Hilbert space of these sensors is typically restricted to a few excitations, we will consider here the general case up to $(N_1,N_2)$-excitations.  For the sake of simplicity, we also consider the particular case of standard photon-counting $\vec \alpha= \vec 0$.
		The computation of the Fisher information in Eq.~\eqref{eq:CFI_PC} requires the use of the joint probability distribution, $p(n_1,n_2|\theta)$, with matrix representation
		%
		%
		%
		%
		\begin{equation}
			p(n_1,n_2|\theta)= 
			\begin{pmatrix}
				p(0,0|\theta) & p(0,1|\theta) & p(0,2|\theta) & \ldots & p(0,N_2|\theta) \\
				p(1,0|\theta) & p(1,1|\theta) & p(1,2|\theta) & \ldots & p(1,N_2|\theta) \\
				p(2,0|\theta) & p(2,1|\theta) & p(2,2|\theta) & \ldots & p(2,N_2|\theta) \\
				\vdots & \vdots & \vdots & \ddots & \vdots \\
				p(N_1,0|\theta) & p(N_1,1|\theta) & p(N_1,2|\theta) & \ldots & p(N_1,N_2|\theta)
			\end{pmatrix}
		\end{equation}
		such that the marginal probability distributions are given by
		\begin{equation}
			p(n_1|\theta)= \sum_{n_2} p(n_1,n_2|\theta) =
			\begin{pmatrix}
				p(0,0|\theta)+p(0,1|\theta)+ \ldots + p(0,N_2|\theta) \\
				p(1,0|\theta)+p(1,1|\theta)+ \ldots + p(1,N_2|\theta)\\
				\vdots \\
				p(N_1,0|\theta)+p(N_1,1|\theta)+ \ldots + p(N_1,N_2|\theta)
			\end{pmatrix},
		\end{equation}
		and equivalently for $p(n_2|\theta)$.
		Consequently, by inserting the definition of $p(n_1,n_2|\theta)$ into Eq.~\eqref{eq:CFI_PC} we obtain
		\begin{equation}
			F_\theta[p(n_1,n_2|\theta)] =
			\frac{ [\partial_\theta p(0,0|\theta)]^2}{p(0,0|\theta)} +	\frac{[\partial_\theta p(0,1|\theta)]^2}{p(0,1|\theta)} 
			+\frac{ [\partial_\theta p(1,0|\theta)]^2}{p(1,0|\theta)}  +
			\frac{[\partial_\theta p(1,1|\theta)]^2}{p(1,1|\theta)}  + \ldots +	\frac{[\partial_\theta p(N_1,N_2|\theta)]^2}{p(N_1,N_2|\theta)} ,
			\label{eq:JointFisher_Ext}
		\end{equation}
		while inserting the expressions for $p(n_1|\theta)$ and $p(n_2|\theta)$ into the marginal Fisher information terms, i.e., $F_\theta[p(n_i|\theta)]$, we get
		\begin{multline}
			F_\theta[p(n_1|\theta)] =
			\frac{[\partial_\theta p(0,0|\theta)+\partial_\theta p(0,1|\theta)+ \ldots + \partial_\theta p(0,N_2|\theta)]^2}{p(0,0|\theta)+p(0,1|\theta)+ \ldots + p(0,N_2|\theta)} \\
			+\ldots + 	\frac{[\partial_\theta p(N_1,0|\theta)+\partial_\theta p(N_1,1|\theta)+ \ldots + \partial_\theta p(N_1,N_2|\theta)]^2}{p(N_1,0|\theta)+p(N_1,1|\theta)+ \ldots + p(N_1,N_2|\theta)},
			\label{eq:MarginalFisher1}
		\end{multline}
		and analogously to $F_\theta[p(n_2|\theta)]$.
		%
		%
		%
		%
		Now, if we rearrange appropriately the terms in Eq.~\eqref{eq:JointFisher_Ext},
		\begin{equation}
			F_\theta[p(n_1,n_2|\theta)] =
			\frac{1}{2}\left(
			\frac{ [\partial_\theta p(0,0|\theta)]^2}{p(0,0|\theta)}+ \ldots +\frac{ [\partial_\theta p(N_1,N_2|\theta)]^2}{p(N_1,N_2|\theta)}
			\right)
			+
			\frac{1}{2}\left(
			\frac{ [\partial_\theta p(0,0|\theta)]^2}{p(0,0|\theta)}+ \ldots +\frac{ [\partial_\theta p(N_1,N_2|\theta)]^2}{p(N_1,N_2|\theta)}
			\right),
		\end{equation}
		and apply the Sedrakyan's inequality, we obtain the following condition
		\begin{multline}
			F_\theta[p(n_1,n_2|\theta)] \geq  
			\frac{1}{2} \left(
			\frac{[\partial_\theta p(0,0|\theta)+ \ldots + \partial_\theta p(0,N_2|\theta)]^2}{p(0,0|\theta)+ \ldots + p(0,N_2|\theta)}
			+\ldots + 	\frac{[\partial_\theta p(N_1,0|\theta)+ \ldots + \partial_\theta p(N_1,N_2|\theta)]^2}{p(N_1,0|\theta)+ \ldots + p(N_1,N_2|\theta)}
			\right)
			\\+
			\frac{1}{2} \left(
			\frac{[\partial_\theta p(0,0|\theta)+ \ldots + \partial_\theta p(N_1,0|\theta)]^2}{p(0,0|\theta)+ \ldots + p(N_1,0|\theta)}
			+\ldots + 	\frac{[\partial_\theta p(0,N_2|\theta)+ \ldots + \partial_\theta p(N_1,N_2|\theta)]^2}{p(N_1,N_2|\theta)+ \ldots + p(N_1,N_2|\theta)}
			\right).
		\end{multline}
		Finally, by recalling the definition of the marginal Fisher information in Eq.~\eqref{eq:MarginalFisher1}, we obtain the bound presented in Eq.~\eqref{eq:uncorrelatedBound}
		\begin{equation}
			F_\theta[p(n_1,n_2|\theta)] \geq \frac{1}{2} \left(F_\theta[p(n_1|\theta)] +F_\theta[p(n_2|\theta)]\right).
		\end{equation}
	\end{widetext}
	
	\let\oldaddcontentsline\addcontentsline
	\renewcommand{\addcontentsline}[3]{}
	\bibliography{Bibliography_AVV}
	\let\addcontentsline\oldaddcontentsline

\end{document}